\documentstyle[graphicx,12pt]{article}
\def\cl{\centerline}
\def\ss{\smallskip}
\def\bs{\bigskip}
\def\ms{\medskip}

\def\beq{\begin{equation}}
\def\eeq{\end{equation}}
\def\ba{\begin{array}}
\def\ea{\end{array}}
\def\bea{\begin{eqnarray}}
\def\eea{\end{eqnarray}}
\def\bc{\begin{center}}
\def\ec{\end{center}}
\def\beqa{\begin{eqnarray}}
\def\eeqa{\end{eqnarray}}
\def\vb#1{\vbox to #1 pt{}}
\def\nn{\nonumber}
\def\ds{\displaystyle}
\def\Eq#1{Eq.~(\ref{#1})}

\def\bold#1{\setbox0=\hbox{$#1$} 
     \kern-.025em\copy0\kern-\wd0 
     \kern.05em\copy0\kern-\wd0 
     \kern-.025em\raise.0433em\box0 } 
\def\ovl{\overline}

\def\lrvpartial{{\partial \! \! \! \! \vbox to 12 pt{}^{\leftrightarrow}}}
\def\lag{{\cal L}}
\def\ket#1{\left\vert #1\right\rangle}

\def\vev#1{\left\langle #1\right\rangle}
\def\ifmath#1{\relax\ifmmode #1\else $#1$\fi}
\def\half{\ifmath{{\textstyle{1 \over 2}}}}
\def\quarter{\ifmath{{\textstyle{1 \over 4}}}}
\def\fourth{\ifmath{{\textstyle{1\over 4}}}}
\def\eighth{\ifmath{{\textstyle{1\over 8}}}}
\def\ra{\rightarrow}
\def\ot{\otimes}
\def\lsim{\raise0.3ex\hbox{$\;<$\kern-0.75em\raise-1.1ex\hbox{$\sim\;$}}}
\def\etal{\hbox{\it et al., }}

\newcommand{\ptmis}{{ {\rm p} \hspace{-0.53 em} \raisebox{-0.27 ex} {/}_T }}

\def\np#1#2#3{           {\sl Nucl. Phys. }{\bf #1} (#2) #3}
\def\pl#1#2#3{           {\sl Phys. Lett. }{\bf #1} (#2) #3}
\def\pr#1#2#3{           {\sl Phys. Rev. }{\bf #1} (#2) #3}
\def\prep#1#2#3{         {\sl Phys. Rep. }{\bf #1} (#2) #3}
\def\prl#1#2#3{          {\sl Phys. Rev. Lett. }{\bf #1} (#2) #3}
\def\epj#1#2#3{          {\sl Eur. Phys. J. }{\bf #1} (#2) #3}

\begin{document}
\thispagestyle{empty}
\begin{titlepage}
\rightline{\hfill hep-ph/9811454}
\rightline{\hfill FISIST/15-98/CFIF}

\begin{center}

{\Large \bf Supersymmetric Theories with R--parity Violation\footnote{
Lectures given at the {\it V Gleb Wataghin School}, Campinas, Brasil, July 1998.}}\\[15mm] 

{{Jorge C. Rom\~ao} 
\hspace{3cm}\\
{\small Departamento de F\'\i sica, Instituto Superior T\'ecnico}\\
%\vspace{-3.5mm}
{\small A. Rovisco Pais, P-1096 Lisboa Codex, Portugal}\\ }

\end{center}
\vspace{5mm}

\begin{abstract} 
In these Lectures we review the Minimal Supersymmetric Standard Model
as well as some of its extensions that include R--Parity
violation. The cases of spontaneous breaking of R--Parity as well as
that of explicit violation through bilinear terms in the
superpotential are studied in detail. The signals at LEP and the
prospects for LHC are discussed.

\end{abstract}

\end{titlepage}

%\tableofcontents

\newpage
\section{The Minimal Supersymmetric Standard Model}

\subsection{Introduction and Motivation}

In recent years it has been established~\cite{LEPI} with great 
precision (in some cases
better than 0.1\%) that the interactions of the gauge bosons with the
fermions are described by the Standard Model (SM)~\cite{SM}. 
However other
sectors of the SM have been tested to a much lesser degree. In fact only now
we are beginning to probe the self--interactions of the gauge bosons through
their pair production at the Tevatron~\cite{Tevatron} and LEP~\cite{LEPII}
and the Higgs sector, responsible for the symmetry breaking has not yet been
tested.

Despite all its successes, the SM still has many unanswered
questions. Among the various candidates to Physics Beyond the Standard
Model, supersymmetric theories play a special role. Although 
there is not yet direct experimental evidence for supersymmetry (SUSY),  
there are many theoretical arguments indicating that SUSY might be of
relevance for physics below the 1 TeV scale. 

\renewcommand{\theenumi}{\it \roman{enumi}}

The most commonly invoked theoretical arguments for SUSY are:
\begin{enumerate}
\item
Interrelates matter fields (leptons and quarks) with force fields
(gauge and/or Higgs bosons).
\item
As local SUSY implies gravity (supergravity) it could provide a way to
unify gravity with the other interactions.
\item
As SUSY and supergravity have fewer divergences than conventional
filed theories, the hope is that it could provide a consistent
(finite) quantum gravity theory.
\item
SUSY can help to understand the mass problem, in particular solve the
naturalness problem ( and in some models even the hierarchy problem)
if SUSY particles have masses $\le {\cal O} (1 \hbox{TeV})$.
\end{enumerate}
\renewcommand{\theenumi}{\bf \Alph{enumi}}
As it is the last argument that makes SUSY particularly attractive for
the experiments being done or proposed for the next decade, let us
explain the idea in more detail. As the SM is not asymptotically free,
at some energy scale $\Lambda$, the interactions must become strong
indicating the existence of new physics. Candidates for this scale
are, for instance, $M_X \simeq {\cal O}(10^{16}\ \hbox{GeV})$ in GUT's
or more fundamentally the Planck scale 
$M_P \simeq {\cal O}(10^{19} \hbox{GeV})$. This alone does not
indicate that the new physics should be related to SUSY, but the
so--called mass problem does. The only consistent way to give masses
to the gauge bosons and fermions is through the Higgs mechanism
involving at least one spin zero Higgs boson. Although the Higgs boson
mass is not fixed by the theory, a value much bigger than $<
H^0>\simeq G_F^{-1/2}\simeq 250\ \hbox{GeV}$ would imply that the Higgs
sector would be strongly coupled making it difficult to understand why
we are seeing an apparently successful perturbation theory at low
energies. Now the one loop radiative corrections to the Higgs boson
mass would give
\beq
\delta m_H^2={\cal O}\left(\frac{\alpha}{4\pi}\right)\ \Lambda^2
\eeq
which would be too large if $\Lambda$ is identified with $\Lambda_{GUT}$
or $\Lambda_{Planck}$. SUSY cures this problem in the following
way. If SUSY were exact, radiative corrections to the scalar masses
squared would be absent because the contribution of fermion loops
exactly cancels against the boson loops. Therefore if SUSY is broken,
as it must, we should have 
\beq
\delta m_H^2={\cal O}\left(\frac{\alpha}{4\pi}\right)\ |m_B^2-m_F^2|
\eeq
We conclude that SUSY provides a solution for the the naturalness
problem if the masses of the superpartners are below ${\cal O}(1\
\hbox{TeV})$. This is the main reason behind all the phenomenological
interest in SUSY.

In the following we will give a brief review of the main aspects of
the SUSY extension of the SM, the so--called Minimal Supersymmetric
Standard Model (MSSM). Almost all the material is covered in many
excellent reviews that exist in the literature~\cite{susy}.

\subsection{SUSY Algebra, Representations and Particle Content}

\subsubsection{SUSY Algebra}

The SUSY generators obey the following algebra
\beqa
\left\{Q_{\alpha},Q_{\beta} \right\} &=&0 \cr
\vb{18}
\left\{\ovl{Q}_{\dot \alpha},\ovl{Q}_{\dot \beta} \right\} &=&0 \cr
\vb{18}
\left\{Q_{\alpha},\ovl{Q}_{\dot \beta} \right\} &=& 
2 \left( \sigma^{\mu} \right)_{\alpha \dot \beta} \, P_{\mu}
\label{susyalgebra}
\eeqa
where 
\beq
\sigma^{\mu}\equiv (1, \sigma^i) \quad ; \quad \ovl{\sigma}^{\mu} \equiv (1
-\sigma^i)
\label{sigma}
\eeq
and $\alpha,\beta,\dot \alpha, \dot \beta=1,2$ (Weyl 2--component spinor
notation). 
The commutation relations with the generators of the Poincar\'e group are
\beqa
\left[P^{\mu},Q_{\alpha} \right] &=&0 \cr
\vb{18}
\left[M^{\mu \nu},{Q}_{\alpha} \right] &=& -i \left( \sigma^{\mu \nu}
\right)_{\alpha}{}^{\beta}\, Q_{\beta}
 \nn
\eeqa
From these relations one can easily derive that the two invariants of
the Poincar\'e group,
\beq
\ba{l}
P^2=P_{\alpha} P^{\alpha} \cr
\vb{18}
W^2=W_{\alpha} W^{\alpha} \cr
\ea
\eeq
where $W^{\mu}$ is the Pauli--Lubanski vector operator
\beq
W_{\mu}=-\frac{i}{2} \epsilon_{\mu \nu \rho \sigma} M^{\nu \rho} P^{\sigma}
\eeq
are no longer invariants of the Super Poincar\'e group. In fact
\beqa
[Q_{\alpha},P^2]=0\cr
\vb{18}
[Q_{\alpha},W^2]\not= 0
\eeqa
showing that the irreducible multiplets will have particles of the
same mass but different spin.

\subsubsection{Simple Results from the Algebra}

From the supersymmetric algebra one can derive two important results:

\begin{enumerate}

\item
{\bf Number of  Bosons = Number of  Fermions}

We have
\beqa
Q_{\alpha} |B> =|F> \quad &;& \quad (-1)^{N_F} |B> =|B> \cr
\vb{18}
Q_{\alpha} |F> =|B> \quad &;& \quad (-1)^{N_F} |F> =- |F> 
\eeqa
where $(-1)^{N_F}$ is the fermion number of a given state. Then we obtain
\beq
Q_{\alpha} (-1)^{N_F}=-(-1)^{N_F} Q_{\alpha} 
\eeq
Using this relation we can show that
\beqa
Tr \left[(-1)^{N_F} \left\{Q_{\alpha},\ovl{Q}_{\dot \alpha} \right\}
\right] 
&\hskip -1mm=\hskip -1mm& 
Tr \left[(-1)^{N_F} Q_{\alpha} \ovl{Q}_{\dot \alpha} +
(-1)^{N_F} \ovl{Q}_{\dot \alpha} Q_{\alpha} \right] \cr
\vb{18}
&\hskip -1mm=\hskip -1mm& 
Tr \left[- Q_{\alpha} (-1)^{N_F} \ovl{Q}_{\dot \alpha} +
Q_{\alpha} (-1)^{N_F} \ovl{Q}_{\dot \alpha} \right] \cr
\vb{18}
&\hskip -1mm=\hskip -1mm& 0 \nn
\eeqa
But using \Eq{susyalgebra} we also have
\beq
Tr \left((-1)^{N_F} \left\{Q_{\alpha},\ovl{Q}_{\dot \alpha} \right\}
\right) =Tr \left((-1)^{N_F} 2 \sigma^{\mu}_{\alpha \dot \alpha} P_{\mu}
\right) 
\eeq
This in turn implies
\ms

\cl{
\fbox{\vb{14}
$
Tr(-1)_{N_F}=\# \hbox{Bosons} -\# \hbox{Fermions}=0
$
}}
\ms

showing that in a given representation the number of degrees of
freedom of the bosons equals those of the fermions.

\item
{\bf $\vev{0|H|0} \ge 0$}

From the algebra we get
\beqa
\left\{Q_1,\ovl{Q}_{\dot 1} \right\}+
\left\{Q_2,\ovl{Q}_{\dot 2} \right\} &=& 2 Tr \left( \sigma^{\mu}\right)
P_{\mu} \cr
\vb{18}
&=& 4 H
\eeqa
Then
\beq
H=\frac{1}{4}\, \left( Q_1 \ovl{Q}_1  + Q_2 \ovl{Q}_{\dot 2 }
+ \ovl{Q}_{\dot 1} Q_1+ \ovl{Q}_{\dot 2} Q_2 \right)
\eeq
and
\beqa
\vev{0|H|0}&\hskip -2mm=\hskip -2mm& 
\left( ||Q_1 \ket{0} ||^2 +||Q_1 \ket{0} ||^2 +
||\ovl{Q}_{\dot 1} \ket{0} ||^2 +||\ovl{Q}_{\dot 2} \ket{0} ||^2 \right)\cr
\vb{18}
&\hskip -2mm\ge\hskip -2mm&0
\eeqa
showing that the energy of the vacuum state is always positive definite.

\end{enumerate}

\subsubsection{SUSY Representations}

We consider separately the massive and the massless case.

\begin{enumerate}
\item
{\bf Massive case}

In the rest frame
\beq
\left\{Q_{\alpha},\ovl{Q}_{\dot \alpha} \right\}= 2\, m\ \delta_{\alpha \dot
\alpha} 
\eeq
This algebra is similar to the algebra of the spin 1/2 creation and
annihilation operators. 
Choose $\ket{\Omega}$ such that
\beq
Q_1 \ket{\Omega}=Q_2 \ket{\Omega}=0
\eeq
Then we have 4 states
\beq
\ket{\Omega} \ ; \ \ovl{Q}_1 \ket{\Omega} \ ; \ 
\ovl{Q}_2 \ket{\Omega} \ ; \ 
\ovl{Q}_1 \ovl{Q}_2 \ket{\Omega} 
\eeq
If $ J_3 \ket{\Omega} =j_3 \ket{\Omega} $ we  show in 
Table~\ref{massive} the values of $J_3$ for the 4 states.
\begin{table}[ht]
\begin{center}
\begin{tabular}{|c|c|}\hline
State& $J_3$ Eigenvalue \cr \hline\hline
$\ket{\Omega}$&$j_3$\cr
$\ovl{Q}_1 \ket{\Omega}$&$j_3 +\frac{1}{2}$\cr
$\ovl{Q}_2 \ket{\Omega}$&$j_3 -\frac{1}{2}$\cr
$\ovl{Q}_1 \ovl{Q}_2 \ket{\Omega}$&$j_3 $\cr\hline
\end{tabular}
\end{center}
\caption{Massive states}
\label{massive}
\end{table}
We notice that there two bosons and two fermions and that the states are
separated by one half unit of spin.

\item 
{\bf Massless case}

If $m=0$ then we can choose $P^{\mu}=(E,0,0,E)$. In this frame
\beq
\left\{Q_{\alpha},\ovl{Q}_{\dot \alpha}\right\} = M_{\alpha \dot \alpha} 
\eeq
where the matrix $M$ takes the form
\beq
M=\left(\matrix{0&0\cr
0&4E\cr}
\right)
\eeq
Then
\beq
\left\{Q_2,\ovl{Q}_2 \right\} = 4 E
\eeq
all others vanish. We have then just {\bf two} states
\beq
\ket{\Omega} \ ; \ \ovl{Q}_2 \ket{\Omega} 
\eeq
If $J_3 \ket{\Omega} =\lambda \ket{\Omega} $ we have the states shown
in Table~\ref{massless},

\bs
\begin{table}[ht]
\begin{center}
\begin{tabular}{|c|c|}\hline
State& $J_3$ Eigenvalue \cr \hline\hline
$\ket{\Omega}$&$\lambda$\cr
$\ovl{Q}_2 \ket{\Omega}$&$\lambda -\frac{1}{2}$\cr\hline
\end{tabular}
\end{center}
\caption{Massless states}
\label{massless}
\end{table}

\end{enumerate}

\subsection{How to Build a SUSY Model}

To construct supersymmetric Lagrangians one normally uses superfield
methods (see for instance~\cite{susy}). In these lectures we do not
have time to go into the details of that construction. Therefore we
will take a more pragmatic view and give the results in the form of a
{\it recipe}. To simplify matters even further we just consider one gauge
group $G$. Then the gauge bosons $W_{\mu}^a$ are in the adjoint 
representation of $G$ and are described by the massless gauge supermultiplet
\beq
V^a\equiv(\lambda^a,W_{\mu}^a)
\eeq
where $\lambda^a$ are the superpartners of the gauge bosons, the
so--called {\it gauginos}. We also consider only one matter chiral
superfield 
\beq
\Phi_i\equiv(A_i,\psi_i) \qquad ; \qquad (i=1,\ldots,N)
\eeq
belonging to some $N$ dimensional representation of $G$. We will give
the rules for the different parts of the Lagrangian for these
superfields. The generalization to the case where we have more
complicated gauge groups and more matter
supermultiplets, like in the MSSM, is straightforward.

\subsubsection{Kinetic Terms}

Like in any gauge theory we have
\beq
{\cal L}_{kin}= -\fourth F_{\mu \nu}^a F^{a \mu \nu} + \frac{i}{2}\
\ovl{\lambda^a} \gamma^{\mu} D_{\mu} \lambda^a 
+ \left(D_{\mu} A\right)^{\dagger} D^{\mu} A + i \ovl{\psi}
\gamma^{\mu} D_{\mu} P_L \psi
\label{Lkin}
\eeq
where the covariant derivative is
\beq
D_{\mu}=\partial_{\mu} + i g W_{\mu}^a T^a
\eeq
In Eq.~(\ref{Lkin}) one should note that $\psi$ is left handed and
that $\lambda$ is a Majorana spinor.

\subsubsection{Self Interactions of the Gauge Multiplet}

For a non Abelian gauge group $G$ we have the usual self--interactions
(cubic and quartic) of the gauge bosons with themselves. These are
well known and we do write them here again. 
But we have a new interaction of the gauge bosons with the
gauginos. In two component spinor notation it reads~\cite{susy}
\beq
{\cal L}_{\lambda \lambda W}=
i g f_{abc}\, \lambda^a \sigma^{\mu} \ovl{\lambda}^b\, W_{\mu}^c + h.c.
\eeq
where $f_{abc}$ are the structure constants of the gauge group $G$ and
the matrices $\sigma^{\mu}$ were introduced in Eq.~(\ref{sigma}).

\subsubsection{Interactions of the Gauge and Matter Multiplets}

In the usual non Abelian gauge theories we have the interactions of
the gauge bosons with the fermions and scalars of the theory. In the
supersymmetric case we
also have interactions of the gauginos with the fermions and scalars
of the chiral matter multiplet. The general form, in two component
spinor notation, is~\cite{susy},
\beqa
{\cal L}_{\Phi W}&=&-g T^a_{ij} W_{\mu}^a \left( \ovl{\psi}_i
\ovl{\sigma}^{\mu} \psi_j + i A_i^* \lrvpartial_{\mu} A_j  \right)
+ g^2 \left(T^a T^b\right)_{ij} W_{\mu}^a W^{\mu b}\, A_i^* A_j \cr
\vb{18}
&&+ ig \sqrt{2}\, T_{ij}^a \left(\lambda^a \psi_j A_i^*
-\ovl{\lambda}^a \ovl{\psi}_i A_j\right)
\eeqa
where the new interactions of the gauginos with the fermions and
scalars are given in the last term.

\subsubsection{Self Interactions of the Matter Multiplet}

These correspond in non supersymmetric gauge theories both to the Yukawa
interactions and to the scalar potential. In supersymmetric gauge theories we
have less freedom to construct these terms. The first step is to
construct the superpotential $W$. This must be a gauge invariant
polynomial function of the {\it scalar} components of the chiral
multiplet $\Phi_i$, that is $A_i$. It {\it does not} depend on
$A_i^*$. In order to have renormalizable theories, the degree of the
polynomial must be at most three. This is in contrast with non
supersymmetric gauge theories where we can construct the scalar
potential with a polynomial up to the fourth degree.

Once we have the superpotential $W$, then the theory is defined and 
the Yukawa interactions are
\beq
{\cal L}_{Yukawa}=-\half \left[
\frac{\partial^2 W}{\partial A_i \partial A_j}\ \psi_i \psi_j +
\left(\frac{\partial^2 W}{\partial A_i \partial A_j}\right)^*\ 
\ovl{\psi}_i \ovl{\psi}_j \right]
\eeq
and the scalar potential is
\beq
V_{scalar}=\half D^a D^a + F_i F_i^*
\eeq
where
\beqa
F_i&=& \frac{\partial W}{\partial A_i}\cr
\vb{18}
D^a&=& g\ A_i^* T_{ij}^a A_j
\eeqa
We see easily from these equations that, if the polynomial degree of W
were higher than three, then the scalar potential would be a polynomial
of degree higher than four and hence non renormalizable.

\subsubsection{Supersymmetry Breaking Lagrangian}

As we have not discovered superpartners of the known particles with
the same mass, we conclude that
 SUSY has to be broken. How this done is the least
understood sector of the theory. In fact, as we shall see, the
majority of the unknown parameters come from this sector. As we do not
want to spoil the good features of SUSY, the form of these SUSY
breaking terms has to obey some restrictions. It has been shown that
the added terms can only be mass terms, or have the same form
of the superpotential, with arbitrary coefficients. These are called
{\it soft terms}. Therefore, for the model that we are considering,
the general form would be\footnote{
We do not consider a term linear in $A$ because we are assuming that
$\Phi$, and hence $A$, are not gauge singlets.}
\beq
{\cal L}_{SB}=m_1^2\, Re(A^2)+m_2^2\, Im(A^2) - m_3\, \left( \lambda^a
\lambda^a + \ovl{\lambda}^a \ovl{\lambda}^a \right) + m_4\, ( A^3 + h.c.)
\eeq
where $A^2$ and $A^3$ are gauge invariant combinations of the scalar
fields. From its form, we see that it only affects the scalar
potential and the masses of the gauginos. The parameters $m_i$
have the dimension of a mass and are in general arbitrary.

\subsubsection{R--Parity}

In many models there is a multiplicatively conserved quantum number
the called {\it R--parity}. It is defined as
\beq
R=(-1)^{2J + 3B +L}
\eeq
With this definition it has the value $+1$ for the known particles and
$-1$ for their superpartners. The MSSM it is a model where R--parity
is conserved. The conservation of R--parity has
three important consequences: {\it i)} SUSY particles are pair
produced, {\it ii)} SUSY particles decay into SUSY particles and {\it iii)}
The lightest SUSY particle is stable (LSP). In Sections 2 and 3 we
will discuss models where R--parity is not conserved.

\subsection{The Minimal Supersymmetric Standard Model}

\subsubsection{The Gauge Group and Particle Content}

We want to describe the supersymmetric version of the SM. Therefore
the gauge group is considered to be that of the SM, that is
\beq
G=SU_c(3)\ot SU_L(2)\ot U_Y(1)
\eeq
We will now describe the minimal particle content. 

\begin{itemize}
\item
{\bf Gauge Fields}

We want to have gauge fields for the gauge group 
$G=SU_c(3)\ot SU_L(2)\ot U_Y(1)$. Therefore we will need three vector
superfields (or vector supermultiplets) $\widehat V_i$ with the following
components:

\beq
\ba{llll}
\widehat V_1\equiv(\lambda',W_1^{\mu})&\ra&U_Y(1)&\cr
\vb{18}
\widehat V_2\equiv(\lambda^a,W_2^{\mu a})&\ra&SU_L(2)\quad ,& a=1,2,3\cr
\vb{18}
\widehat V_3\equiv(\widetilde{g}^b,W_3^{\mu b})&\ra&SU_c(3)\quad ,&b=1,\ldots,8
\ea
\eeq
where $W_i^{\mu}$ are the gauge fields and $\lambda',\lambda$ and
$\widetilde{g}$ are the $U_Y(1)$ and $SU_L(2)$ gauginos and the
gluino, respectively.

\item
{\bf Leptons}

The leptons are described by chiral supermultiplets. As each chiral
multiplet only describes one helicity state, we will need two chiral
multiplets for each charged lepton\footnote{We will assume that the
neutrinos do not have mass.}.
\begin{table}[ht]
\vspace{-5mm}
\begin{center}
\begin{tabular}{|l|c|}\hline
Supermultiplet&$SU_c(3)\ot SU_L(2)\ot U_Y(1)$\cr
&Quantum Numbers\cr\hline\hline
\vb{14}
$\widehat L_i\equiv(\widetilde{L},L)_i$&$(1,2,-\half)$\cr
\vb{14}
$\widehat R_i\equiv(\widetilde{\ell}_R,\ell^c_L)_i$&$(1,1,1)$\cr\hline
\end{tabular}
\end{center}
\vspace{-5mm}
\caption{Lepton Supermultiplets}
\label{table:leptons}
\end{table}
The multiplets are given in Table~\ref{table:leptons}, where 
the $U_Y(1)$ hypercharge is defined through $Q=T_3+Y$. Notice that
each helicity state corresponds to a complex scalar and that $\hat
L_i$ is a doublet of $SU_L(2)$, that is
\beq
\widetilde{L}_i=\left(
\ba{l}
\widetilde{\nu_i}_L\cr
\vb{16}
\widetilde{\ell}_{iL}
\ea
\right)
\qquad ; \qquad
L_i=
\left(
\ba{l}
\nu_{iL}\cr
\ell_{iL}
\ea
\right)
\eeq

\item
{\bf Quarks}

The quark supermultiplets are given in Table~\ref{table:quarks}. The
supermultiplet $\widehat Q_i$ is also a  doublet of $SU_L(2)$, that
is
\begin{table}[ht]
\vspace{-5mm}
\begin{center}
\begin{tabular}{|l|c|}\hline
Supermultiplet&$SU_c(3)\ot SU_L(2)\ot U_Y(1)$\cr
&Quantum Numbers\cr\hline\hline
\vb{14}
$\widehat Q_i\equiv(\widetilde{Q},Q)_i$&$(3,2,\frac{1}{6})$\cr
\vb{14}
$\widehat D_i\equiv(\widetilde{d}_R,d^c_L)_i$&$(3,1,\frac{1}{3})$\cr
\vb{14}
$\widehat U_i\equiv(\widetilde{u}_R,u^c_L)_i$&$(3,1,-\frac{2}{3})$\cr\hline
\end{tabular}
\end{center}
\vspace{-5mm}
\caption{Quark Supermultiplets}
\label{table:quarks}
\end{table}
\beq
\widetilde{Q}_i=\left(
\ba{l}
\widetilde{u}_{iL}\cr
\vb{16}
\widetilde{d}_{iL}
\ea
\right)
\qquad ; \qquad
Q_i=
\left(
\ba{l}
u_{iL}\cr
d_{iL}
\ea
\right)
\eeq

\item
{\bf Higgs Bosons}

Finally the Higgs sector. In the MSSM we need at least two Higgs
doublets. This is in contrast with the SM where only one Higgs doublet
is enough to give masses to all the particles. The reason can be explained in
two ways. Either the need to cancel the anomalies, or the fact that,
due to the analyticity of the superpotential, we have to have two
Higgs doublets of opposite hypercharges to give masses to the up and
down type of quarks. The two supermultiplets, with their quantum
numbers, are given in Table~\ref{table:higgs}.
\begin{table}[ht]
\vspace{-5mm}
\begin{center}
\begin{tabular}{|l|c|}\hline
Supermultiplet&$SU_c(3)\ot SU_L(2)\ot U_Y(1)$\cr
&Quantum Numbers\cr\hline\hline
\vb{14}
$\widehat H_1\equiv(H_1,\widetilde{H}_1)$&$(1,2,-\frac{1}{2})$\cr
\vb{14}
$\widehat H_2\equiv(H_2,\widetilde{H}_2)$&$(1,2,+\frac{1}{2})$\cr\hline
\end{tabular}
\end{center}
\vspace{-5mm}
\caption{Higgs Supermultiplets}
\label{table:higgs}
\end{table}

\end{itemize}

\subsubsection{The Superpotential and SUSY Breaking Lagrangian}

The MSSM Lagrangian is specified by the R--parity conserving 
superpotential $W$ 
\beqa
W&=&\varepsilon_{ab}\left[ 
 h_U^{ij}\widehat Q_i^a\widehat U_j\widehat H_2^b 
+h_D^{ij}\widehat Q_i^b\widehat D_j\widehat H_1^a 
+h_E^{ij}\widehat L_i^b\widehat R_j\widehat H_1^a 
-\mu\widehat H_1^a\widehat H_2^b \right] \
\eeqa
where $i,j=1,2,3$ are generation indices, $a,b=1,2$ are $SU(2)$ 
indices, and $\varepsilon$ is a completely antisymmetric $2\times2$  
matrix, with $\varepsilon_{12}=1$. The coupling matrices $h_U,h_D$ and
$h_E$ will give rise to the usual Yukawa interactions needed to give
masses to the leptons and quarks. If it were not for the need to break
SUSY, the number of parameters involved would be less than in the
SM. This can be seen in Table~\ref{SMvsMSSM}.

The most general SUSY soft breaking is 
\begin{eqnarray} 
V_{SB}&=& 
M_Q^{ij2}\widetilde Q^{a*}_i\widetilde Q^a_j+M_U^{ij2} 
\widetilde U_i\widetilde U^*_j+M_D^{ij2}\widetilde D_i 
\widetilde D^*_j
+M_L^{ij2}\widetilde L^{a*}_i\widetilde L^a_j 
+M_R^{ij2}\widetilde R_i\widetilde R^*_j
+m_{H_1}^2 H^{a*}_1 H^a_1\cr
\vb{18}
&&+m_{H_2}^2 H^{a*}_2 H^a_2 
+ \left[\half M_s\lambda_s\lambda_s+\half M\lambda\lambda 
+\half M'\lambda'\lambda'+h.c.\right]\cr
\vb{18}
&&
+\varepsilon_{ab}\left[ 
A_U^{ij}h_U^{ij}\widetilde Q_i^a\widetilde U_j H_2^b 
+A_D^{ij}h_D^{ij}\widetilde Q_i^b\widetilde D_j H_1^a 
+A_E^{ij}h_E^{ij}\widetilde L_i^b\widetilde R_j H_1^a 
-B\mu H_1^a H_2^b \right] 
\end{eqnarray} 
\begin{table}[ht]
\begin{center}
\begin{tabular}{|c|c|c|c|}\hline
Theory&Gauge&Fermion& Higgs\cr
&Sector&Sector& Sector\cr\hline\hline
SM&$e,g,\alpha_s$&$h_U,h_D,h_E$&$\mu^2,\lambda$\cr\hline
MSSM&$e,g,\alpha_s$&$h_U,h_D,h_E$&$\mu$\cr\hline
Broken
MSSM&$e,g,\alpha_s$&$h_U,h_D,h_E$&$\mu,M_1,M_2,M_3,A_U,A_D,A_E,B$\cr
&&&$m_{H_2}^2,m_{H_1}^2,m_Q^2,m_U^2,m_D^2,m_L^2,m_R^2$\cr\hline
\end{tabular}
\end{center}
\caption{Comparative counting of parameters}
\label{SMvsMSSM}
\end{table}

\subsubsection{Symmetry Breaking}

The electroweak symmetry is broken when the two Higgs doublets  
$H_1$ and $H_2$ acquire VEVs
\beq
H_1={{{1\over{\sqrt{2}}}[\sigma^0_1+v_1+i\varphi^0_1]}\choose{ 
H^-_1}}\,,\qquad 
H_2={{H^+_2}\choose{{1\over{\sqrt{2}}}[\sigma^0_2+v_2+ 
i\varphi^0_2]}} 
\eeq
with $m_W^2=\quarter g^2v^2$ and $v^2\equiv v_1^2+v_2^2=(246 \;
\rm{GeV})^2$. The full scalar potential at tree level is 
\beq
V_{total}  = \sum_i \left| { \partial W \over \partial z_i} \right|^2 
	+ V_D + V_{soft} 
\eeq
The scalar potential contains linear terms 
\beq
V_{linear}=t_1^0\sigma^0_1+t_2^0\sigma^0_2
\eeq
where
\begin{eqnarray} 
t_1&=&(m_{H_1}^2+\mu^2)v_1-B\mu v_2+
\eighth(g^2+g'^2)v_1(v_1^2-v_2^2)\,, 
\cr
\vb{18}
t_2&=&(m_{H_2}^2+\mu^2)v_2-B\mu v_1- 
\eighth(g^2+g'^2)v_2(v_1^2-v_2^2) 
\end{eqnarray} 
The minimum of the potential occurs for $t_i=0$ ($i=1,2$). One can
easily see that this occurs for $m_{H_2}^2 < 0$.

\subsubsection{The Fermion Sector}

The charged gauginos mix with the charged higgsinos giving the
so--called charginos. In a basis where  
$\psi^{+T}=(-i\lambda^+,\widetilde H_2^+)$ 
and $\psi^{-T}=(-i\lambda^-,\widetilde H_1^-)$, the chargino 
mass terms in the Lagrangian are 
\begin{equation} 
{\cal L}_m=-{1\over 2}(\psi^{+T},\psi^{-T}) 
\left(\matrix{{\bold 0} & \bold M_C^T \cr {\bold M_C} &  
{\bold 0} }\right) 
\left(\matrix{\psi^+ \cr \psi^-}\right)+h.c. 
\label{eq:chFmterm} 
\end{equation} 
where the chargino mass matrix is given by 
\begin{equation} 
{\bold M_C}=\left[\matrix{ 
M_2 & {\textstyle{1\over{\sqrt{2}}}}gv_2 \cr 
{\textstyle{1\over{\sqrt{2}}}}gv_1 & \mu }
\right]
\label{eq:ChaM6x6} 
\end{equation} 
and $M$ is the $SU(2)$ gaugino soft mass. 
The chargino mass matrix is diagonalized by two rotation matrices 
$\bold U$ and $\bold V$ defined by
\beq
F_i^-=\bold{U}_{ij}\, \psi_j^- \quad ; \quad F_i^+=\bold{V}_{ij}\, \psi_j^+
\eeq
Then 
\beq
\bold{U}^* \bold{M_C} \bold{V}^{-1} = \bold{M_{CD}}
\eeq
where $\bold{M_{CD}}$ is the diagonal chargino mass matrix.
To determine $\bold{U}$ and $\bold{V}$ we note that
\beq
\bold{M^2_{CD}}=\bold{V} \bold{M_C^{\dagger}} \bold{M_C} \bold{V}^{-1} =
\bold{U^*} \bold{M_C} \bold{M_C^{\dagger}} (\bold{U^*})^{-1}
\eeq
implying that $\bold{V}$ diagonalizes $\bold{M_C^{\dagger} M_C}$ and 
$\bold{U^*}$ diagonalizes $\bold{M_C M_C^{\dagger}}$. 
In the previous expressions the $F_i^{\pm}$ are two component
spinors. We construct the four component Dirac spinors out of the two
component spinors with the
conventions\footnote{Here we depart from the conventions of
ref.~\cite{susy} because we want the $\chi^{-}$ to be the particle and
not the anti--particle.},
\beq
\chi_i^-=\left(\matrix{
F_i^- \cr
\vb{16}
\overline{F_i^+}\cr}
\right)
\eeq

In the basis $\psi^{0T}= 
(-i\lambda',-i\lambda^3,\widetilde{H}_1^1,\widetilde{H}_2^2)$ 
the neutral fermions mass terms in the Lagrangian are given by 
\begin{equation} 
{\cal L}_m=-\frac 12(\psi^0)^T{\bold M}_N\psi^0+h.c.   
\label{eq:NeuMLag} 
\end{equation} 
where the neutralino mass matrix is 
\begin{equation} 
{\bold M}_N=\left[  
\begin{array}{cccc}  
M_1 & 0 & -\frac 12g^{\prime }v_1 & \frac 12g^{\prime }v_2 \\   
0 & M_2 & \frac 12gv_1 & -\frac 12gv_2 \\   
-\frac 12g^{\prime }v_1 & \frac 12gv_1 & 0 & -\mu  \\   
\frac 12g^{\prime }v_2 & -\frac 12gv_2 & -\mu  & 0 \\   
\end{array}  
\right] 
\label{eq:NeuM4x4} 
\end{equation} 
and $M_1$ is the $U(1)$ gaugino soft mass. This neutralino mass  
matrix is diagonalized by a $4\times 4$ rotation matrix $\bold N$ such that 
\begin{equation} 
{\bold N}^*{\bold M}_N{\bold N}^{-1}={\rm diag}(m_{F^0_1},m_{F^0_2}, 
m_{F^0_3},m_{F^0_4}) 
\label{eq:NeuMdiag} 
\end{equation} 
and
\beq
F^0_k=\bold{N}_{kj}\, \psi^0_j
\eeq
The four component Majorana neutral fermions are obtained from the
two component via the relation
\beq
\chi_i^0=\left(\matrix{
F_i^0\cr
\vb{16}
\overline{F_i^0}\cr}
\right)
\eeq

\subsubsection{The Higgs Sector}

In the MSSM there are charged and neutral Higgs bosons. Here we just
discuss the neutral Higgs bosons. Some discussion on charged Higgs
bosons is included in Section~\ref{ChargedScalars}. For a complete
discussion see ref.~\cite{susy}. In the neutral Higgs sector we have two
complex scalars that correspond to four real neutral fields. If the
parameters are real (CP is conserved in this sector) the real and
imaginary parts do not mix and we get two CP--even and two CP--odd neutral
scalars. The form of the mass matrices can be very much affected by
the large radiative corrections due to top--stop loops and we will
discuss both cases separately.

\subsubsection*{Tree Level}

The tree level mass matrices are

\beq
\bold{M_R}^2=\left(
\matrix{
\cot \beta & -1\cr
\vb{16}
-1 & \tan \beta\cr}\right)\,
\half m_Z^2 \sin 2 \beta \ + \
\left( \matrix{
\tan \beta & -1 \cr
\vb{16}
-1 &\cot \beta\cr}\right)\,
\Delta
\eeq
and
\beq
\bold{M_I}^2=\left(
\matrix{
\tan \beta & -1 \cr
\vb{16}
-1 &\cot \beta\cr}\right)\,
\Delta \hskip 1cm \hbox{where} \hskip 1cm \Delta=B\mu
\eeq
Notice that $det (\bold{M_I}^2)=0$. In fact the eigenvalues of
$\bold{M_I}^2$ are $0$ and $m_A^2=2\Delta / \sin 2 \beta$. The zero mass
eigenstate is the Goldstone boson to be {\it eaten} by the $Z^0$. $A$
is the remaining pseudo--scalar. For the real part we have two
physical states, $h$ and $H$, with masses
\beq
m_{h,H}^2=\half\left[m_A^2 +m_Z^2 \mp 
\sqrt{(m_A^2+m_Z^2)^2 -4 m_A^2 m_Z^2 \cos^2 2 \beta}\right]
\eeq
with the {\it tree level} relation
\beq
m_h^2+m_H^2=m_A^2+m_Z^2
\eeq
which implies
\beqa
&&m_h<m_A<m_H\cr
\vb{16}
&&m_h<m_Z<m_H
\label{MassRelations}
\eeqa

\subsubsection*{Radiative Corrections}

The mass relations in Eq.~(\ref{MassRelations}) were true before it
was clear that the top mass is very large. The radiative corrections
due to the top mass are in fact quite large and can not be neglected
if we want to have a correct prediction. The whole picture is quite
complicated\cite{RCHiggsMass}, but here we just give the biggest
correction due to top--stop loops. The mass matrices are now, in this
approximation,
\beqa
\bold{M_R}^2&=&\left(
\matrix{
\cot \beta & -1\cr
\vb{16}
-1 & \tan \beta\cr}\right)\,
\half m_Z^2 \sin 2 \beta \ + \
\left( \matrix{
\tan \beta & -1 \cr
\vb{16}
-1 &\cot \beta\cr}\right)\,
\Delta \cr
\vb{24}
&&+ \frac{3g^2}{16\pi^2m_W^2}\, 
\left( \matrix{
\Delta_{11} & \Delta_{12} \cr
\vb{16}
\Delta_{21} & \Delta_{22}\cr}\right)\,
\eeqa
and
\beq
\bold{M_I}^2=\left(
\matrix{
\tan \beta & -1 \cr
\vb{16}
-1 &\cot \beta\cr}\right)\,
\Delta 
\eeq
where 
\beq
\Delta=B\mu-\frac{3g^2}{64\pi^2\sin^2\beta}\, \frac{m_t^2}{m_W^2}\,
\frac{A\mu}{m^2_{\tilde{t}_1}-m^2_{\tilde{t}_2}}
\left[f(m^2_{\tilde{t}_1})-f(m^2_{\tilde{t}_2})\right]
\eeq
with
\beq
f(m^2)=2 m^2 \left[\log\left(\frac{m^2}{Q^2}\right) -1 \right]
\eeq
The $\Delta_{ij}$ are complicated expressions\cite{RCHiggsMass}. The
most important is
\beq
\Delta_{22}=\frac{m_t^4}{\sin^2 \beta}\,
\log \left(\frac{m^2_{\tilde{t}_1}m^2_{\tilde{t}_2}}{m_t^4}\right)
\label{Delta22}
\eeq
Due to the strong dependence on the top mass in Eq.~(\ref{Delta22})
the CP--even states are the most affected. The mass of the lightest
Higgs boson, $h$ can now be as large as $140\ GeV$\cite{RCHiggsMass}.

\subsection{The Constrained Minimal Supersymmetric Standard Model}

We have seen in the previous section that the parameters of the MSSM
can be considered arbitrary at the weak scale. This is completely
consistent. However the number of independent parameters in
Table~\ref{SMvsMSSM} can be
reduced if we impose some further constraints. That is usually done by
embedding the MSSM in a grand unified scenario. Different schemes are
possible but in all of them some kind of unification is imposed at the
GUT scale. Then we run the Renormalization Group (RG) equations down
to the weak scale to get the values of the parameters at that
scale. This is sometimes called the constrained MSSM model.

Among the possible scenarios, the most popular is the MSSM coupled to
$N=1$ Supergravity (SUGRA). 
Here at $M_{GUT}$ one usually takes the conditions:
\bea
&&A_t = A_b = A_{\tau} \equiv A \:, B=A-1 \:, \cr
&&\vb{18}
m_{H_1}^2 = m_{H_2}^2 = M_{L}^2 = M_{R}^2 = m_0^2 \:,
M_{Q}^2 =M_{U}^2 = M_{D}^2 = m_0^2 \:, \cr
&&\vb{18}
M_3 = M_2 = M_1 = M_{1/2}
\eea

The counting of free parameters\footnote{For one family and without counting
the gauge couplings.} is done in Table~\ref{table:SUGRA}. 

\begin{table}[ht]
\begin{center}
\begin{tabular}{|c|c|c|}\hline
Parameters & Conditions & Free Parameters \cr \hline\hline
$h_t$, $h_b$, $h_{\tau}$, $v_1$, $v_2$
&$m_W$, $m_t$, $m_b$, $m_{\tau}$ & $\tan
\beta$ \cr \hline
$A$, $m_0$, $M_{1/2}$, $\mu$&$t_i=0$, $i=1,2$& 2 Extra free
parameters\cr \hline
Total = 9&Total = 6 &Total = 3\cr\hline
\end{tabular}
\end{center}
\caption{\small
Counting of free parameters in the MSSM coupled to N=1 SUGRA
}
\label{table:SUGRA}
\end{table}

It is remarkable that with so few parameters we can get
the correct values for the parameters,
in particular $m_{H_2}^2 <0$. For this to happen the top Yukawa
coupling has to be large which we know to be true.

\newpage

\renewcommand{\theenumi}{\arabic{enumi}}

\section{Spontaneous Breaking of R--parity}

\subsection{Introduction}

Most studies of supersymmetric phenomenology have been made
in the framework of the MSSM  which assumes the conservation of a discrete 
symmetry called R--parity ($R_p$) as has been explained in the previous
Section. Under this symmetry all the standard
model particles are R-even, while their superpartners are R-odd.
$R_p$ is related to the spin (S), total lepton (L), and
baryon (B)  number according to $R_p=(-1)^{(3B+L+2S)}$. 
Therefore the requirement of baryon and lepton number  conservation 
implies the conservation of $R_p$. Under this assumption the
SUSY particles must be pair-produced,
every SUSY particle decays into another SUSY particle and
the lightest of them is absolutely stable. These three
features underlie all the experimental searches for
new supersymmetric states.

However, neither gauge invariance nor SUSY require $R_p$ 
conservation. The most general supersymmetric extension of the
standard model contains explicit $R_p$ violating interactions that
are consistent with both gauge invariance and supersymmetry. 
Detailed analysis of the constraints on these models and 
their possible signals have been made\cite{Barger89}.
In general, there are too many independent couplings and some of these
couplings have to be  set to zero to avoid the proton to decay too fast. 

For these reasons we restrict, in this Section, 
our attention to the possibility
that $R_p$ can be an exact symmetry of the Lagrangian,
broken spontaneously through the Higgs 
mechanism\cite{Majoron,Masiero90,Nogueira90}.
This may occur via  nonzero
vacuum expectation values for scalar neutrinos, such as
\begin{equation}
v_R = \vev {\tilde{\nu}_{R\tau}}\qquad ;
\qquad v_L = \vev {\tilde{\nu}_{L\tau}}\ .
\end{equation}
If spontaneous $R_p$ violation occurs
in absence of any additional gauge symmetry, it leads to the
existence of a physical massless Nambu-Goldstone boson, called
Majoron (J)\cite{Majoron}. In these models there is a
new decay mode for the $Z^0$ boson, $Z^0 \ra \rho + J $, 
where $\rho$ is a light scalar. This decay mode
would increase the invisible $Z^0$ width by an amount 
equivalent to $1/2$ of a light neutrino family. 
The LEP measurement on the number of such 
neutrinos\cite{LEPI} is enough to exclude any model where the
Majoron is not mainly an isosinglet\cite{Concha90}. The simplest
way to avoid this limit is to extend the MSSM,
so that the $R_p$ breaking is driven by isosinglet
VEVs, so that the Majoron is mainly a singlet\cite{Masiero90}. In this
section we will describe in detail this model for Spontaneously Broken
R--Parity (SBRP) and compare its predictions with the
experimental results.

\subsection{A Viable Model for Spontaneous R--parity Breaking}

In order to set up our notation we recall the basic ingredients
of the model for spontaneous violation of R parity and
lepton number proposed in\cite{Masiero90}. 
The superpotential is given by
\begin{eqnarray}
\label{W_RP}
W=& &h_u  Q H_u U + h_d  H_d Q D + h_e L H_d R 
\cr
\vb{20}
&+& (h_0 H_u H_d - \varepsilon^2 ) \Phi \cr
\vb{20}
&+& h_{\nu} L H_u \nu^c +  h \Phi  S \nu^c
\end{eqnarray}
This superpotential conserves {\sl total} lepton number and 
$R_p$. The superfields $(\Phi ,{\nu^c}_i,S_i)$ are singlets under
$SU_{2} \ot U(1)$ and carry a conserved lepton number assigned 
as $(0,-1,1)$ respectively. All couplings 
$h_u,h_d,h_e,h_{\nu},h$ 
are described by arbitrary matrices in generation space which 
explicitly break flavor conservation.

As we will show in the next section these singlets may 
drive the spontaneous violation of $R_p$\cite{Masiero90,pot3} 
leading to the existence of a Majoron given by the imaginary part 
of
\begin{equation}
{v_L^2 \over Vv^2} (v_u H_u - v_d H_d) +
	      {v_L \over V} \tilde{\nu_{\tau}} -
	      {v_R \over V} \tilde{\nu^c}_{\tau} +
	      {v_S \over V} \tilde{S_{\tau}} 
\end{equation}
where the isosinglet VEVs
\begin{equation}
v_R = \vev {\tilde{\nu}_{\tau}^c}\qquad , \qquad
v_S = \vev {\tilde{S_{\tau}}} 
\end{equation}
with $V = \sqrt{v_R^2 + v_S^2}$, characterize $R_p$ or lepton
number breaking and the isodoublet VEVs
\begin{equation}
v_u = \vev {H_u} \quad , \quad v_d = \vev {H_d} 
\quad , \quad v_L = \vev {\tilde{\nu}_{L\tau}}
\end{equation}
drive electroweak breaking and the fermion masses.

\subsection{Symmetry Breaking}

\subsubsection{Tree Level Breaking}

First we are going to show that the scalar potential has 
vacuum solutions that break $R_p$. Contrary to the case of the MSSM
described in the previous section, the model described by 
\Eq{W_RP} can achieve the breaking of $SU(2) \times U(1)$ at tree
level, without the need of having some negative mass squared driven by
some RG equation. The complete model has three generations and, as we will
see, some mixing among generations is needed for consistency.
But for the analysis of the scalar potential we are going 
to consider, for simplicity, a 1-generation model.

Before we write  down the scalar potential we need to specify the 
soft breaking terms. We write them in the form given in the 
spontaneously broken $N=1$ supergravity models, that is
\begin{eqnarray}
\label{V_SOFT}
V_{soft}=&&\tilde{m}_0 \left[ -A h_0 \Phi H_u H_d -B 
\varepsilon^2 \Phi + C h_{\nu} \tilde{\nu}^c \tilde{\nu} H_u 
+D h \Phi \tilde{\nu}^c S + h. c. \right] \cr
\vb{20}
&+&\tilde{m}_u^2 |H_u |^2 + \tilde{m}_d^2 |H_d |^2 
+ \tilde{m}_L^2 |\tilde{\nu}|^2 + \tilde{m}_R^2 |\tilde{\nu}^c|^2
+\tilde{m}_S^2 |S|^2 + \tilde{m}_F^2 |\Phi |^2 
\end{eqnarray}
At unification scale we have
\begin{eqnarray}
&C=D=A \qquad ; \qquad B=A-2& \cr
\vb{20}
&\tilde{m}_u^2=\tilde{m}_d^2= \cdots = \tilde{m}_0^2 &
\end{eqnarray}
At low energy these relations will be modified by the 
renormalization group  evolution. For simplicity we take 
$C=D=A$ and $B=A-2$ but let\footnote{ Notice that for 
$\vev{H_u}\not=\vev{H_d}$ we must have $\tilde{m}_u^2 
\not=\tilde{m}_d^2$ even in MSSM.}
$\tilde{m}_u^2\not=\tilde{m}_d^2 \not= \cdots \not=\tilde{m}_0^2$.
Then the neutral scalar potential is given by
\begin{eqnarray}
V_{total}=&& {1 \over 8}\ (g^2+g'^2) \left[ |H_u |^2 - |H_d |^2 - 
|\tilde{\nu}|^2 \right]^2 \cr
\vb{20}
&+&| h \Phi S + h_{\nu} \tilde{\nu} H_u |^2 + | h_0 \Phi H_u |^2 + 
|h 
\Phi \tilde{\nu}^c|^2 \cr
\vb{20}
&+&|h \Phi \tilde{\nu}^c|^2 +| -h_0 \Phi H_d + h_{\nu} \tilde{\nu} 
\tilde{\nu}^c |^2 + |-h_0 H_u H_d + h \tilde{\nu}^c S -\varepsilon^2 
|^2 \cr
\vb{20}
&+&\tilde{m}_0\left[ -A\left(-h\Phi \tilde{\nu}^c S + h_0 \Phi H_u 
H_d -h_{\nu} \tilde{\nu} H_u \tilde{\nu}^c \right) + (2-A) 
\varepsilon^2 \Phi + h.c. \right] \cr
\vb{20}
&+& \sum_i \tilde{m}_i^2 |z_i|^2 
\end{eqnarray}
where $z_i$ stand for any of the neutral scalar fields. The 
stationary equations are then
\beq
\label{min}
\left.{\partial V \over \partial z_i}\right\vert_{z_i=v_i}=0\ .
\eeq
These are a set of six nonlinear equations that should be 
solved for the VEVs for each set of parameters. To understand 
the problems in solving these equations we just right down one of 
them, for instance
\bea
\left.{\partial V \over \partial H_d}\right\vert_{H_d=v_d}=
&-&\left[ {1 \over 4} (g^2+g'^2) ( v_u^2-v_d^2-v_L^2)-h_0^2 v_u^2 
-\tilde{m}_d^2-h_0 v_F \right] v_d \cr
\vb{20}
&-& (A h_0 \tilde{m}_0 v_F + h h_0 v_R v_S -h_0 \varepsilon^2 ) 
v_u - h_{\nu} v_L v_R h_0 v_F =0 
\eea

 Also it is important to realize that it is not enough to find a 
solution of these equations but it is necessary to show that it is a 
minimum of the potential. To find the solutions we did not directly 
solve \Eq{min} but rather use the following three step 
procedure:

\begin{enumerate}

\item
{\sl Finding solutions of the extremum equations}

We start by taking random values for 
$h$, $h_0$, $h_{\nu}$, $A$, $\varepsilon^2$, $\tilde{m}_0$, 
$v_R$, and $v_S$.
Then choose $\tan \beta=v_u/v_d$ and fix 
$v_u,v_d$ by 
\beq
m_W^2=\half g^2 (v_u^2+v_d^2+v_L^2)
\eeq
Finally we solve the extremum equations {\sl exactly} for 
$\tilde{m}_u^2$, $\tilde{m}_d^2$, $\ldots$, $\tilde{m}_0^2 $. This is 
possible because they are linear equations on the mass squared 
terms.

\item
{\sl  Showing that the solution is a minimum}

To show that the solution is a true minimum we 
calculate the squared mass matrices. These are
\bea
\label{mm}
&{M_R^2}_{ij}=\left[ \half \left( {\ds \partial^2 V \over \ds \partial z_i 
\partial z_j } + c.c. \right) 
+   {\ds \partial^2 V \over\ds \partial z_i \partial 
z_j^* } \right]_{z_i=v_i}\cr
\vb{20}
&{M_I^2}_{ij}=\left[ -\half \left( {\ds \partial^2 V \over \ds \partial z_i 
\partial z_j } + c.c. \right) 
+   {\ds \partial^2 V \over\ds \partial z_i \partial 
z_j^* } \right]_{z_i=v_i} 
\eea
The solution is a minimum if all nonzero eigenvalues
are positive. A consistency check is that we should get two zero 
eigenvalues for $M_I^2$ corresponding to the Goldstone boson of 
the $Z^0$ and to the majoron $J$.
\ss

\item
{\sl Comparing with other minima}

There are three kinds of minima to which we should compare our 
solution. 
\bea
&\bullet& v_u=v_d=v_L=v_R=v_S=0 \quad ; \quad v_F\not=0 \cr
\vb{20}
&\bullet&v_L=v_R=v_S=0 \quad ; \quad
v_u,v_d,v_F\not=0 \cr
\vb{20}
&\bullet&v_u=v_d=v_L=0 \quad ; \quad
v_R,v_S,v_F\not=0 
\eea

As a final result we found a large region in 
parameter space where our solution that breaks $R_p$ and
$SU_{2} \ot U(1)$ is an absolute minimum.

\end{enumerate}

\subsubsection{Radiative Breaking}

We tried to constrain the model of \Eq{W_RP} by imposing boundary
conditions at some unification scale and using the RG equations to
evolve the parameters to the weak scale. Despite all our efforts we
were not able to obtain radiative spontaneous breaking of {\sl both} Gauge
Symmetry {\sl and} $R_p$ in this simplest model.

To show the point that this could be achieved, we consider instead
a model with Rank--4 unification, given by
the following superpotential:
\bea
\label{W_Ara}
W&=& h_u u^c Q H_u + h_d d^c Q H_d + h_e e^c L H_d   \cr
&&\vb{20}
+h_0 H_u H_d \Phi +
h_{\nu} \nu^c L H_u + 
h \Phi \nu^c S +
\lambda \Phi^3
\eea
The boundary conditions at unification are
\bea
\label{universality}
&&A_u = A = A_0 = A_{\nu} = A_{\lambda} \:, \cr
&&\vb{20}
M_{H_u}^2 = M_{H_d}^2 = M_{\nu_L}^2 = M_{u^c}^2 = M_{Q}^2 =m_0^2 \:,
\cr
&&\vb{20}
M_{\nu^c}^2 = C_{\nu^c}  m_0^2 \ ;M_{S}^2 = C_{S}  m_0^2 \ ;
M_{\Phi}^2 = C_{\Phi}  m_0^2  \:,  \cr
&&\vb{20}
M_3 = M_2 = M_1 = M_{1/2} 
\eea
We run the RGE from the unification scale 
$M_U \sim 10^{16}$ GeV down to the weak scale. In doing 
this we randomly give values at the unification scale.
After running the RGE we have a complete set of parameters, 
Yukawa couplings and soft-breaking masses $m^2_i(RGE)$ 
to study the minimization of the potential,
\beq
V_{total}  = \sum_i \left| { \partial W \over \partial z_i} \right|^2
	+ V_D + V_{SB} + V_{RC}
\eeq
To solve the extremum equations we use the method described before:

\begin{enumerate}

\item
The value of $v_u$ is determined from $m_{top}=h_t v_u$ for
$m_{top}=175 \pm 5$ GeV. If $v_u$ determined in this way is too high 
we go back to the RGE and choose another starting point.
\item
$v_d$ and $\tan(\beta)$ are then determined by $m_W$.
\item
$v_L$ is obtained by solving approximately the corresponding extremum
equation.
\item
We then vary randomly $m_0$, $v_R$, $v_S$, $v_{\phi}$.
\item
We solve the extremum equations for the soft breaking masses, 
which we now call $m^2_i$. 
\item
Calculate numerically the eigenvalues to make sure it is a minimum.

\end{enumerate}

After doing this we end up with a set of points for which: {\it i)} 
The Yukawa couplings and the gaugino mass terms are given by the RGE's,
{\it ii)}
For a given set of $m^2_i$ each point is also a solution of the minimization
of the potential that breaks $R_p$,
{\it iii)}
However,  the $m^2_i$ obtained by minimizing
the potential differ from those obtained from the RGE, $m^2_i(RGE)$. 
Our  goal is to find solutions that obey
\beq
m^2_i=m^2_i(RGE) \quad \forall i
\eeq
To do that we define a function
\beq
\label{eta_def}
\eta= max \left( \frac{m^2_i}{m^2_i(RGE)},\frac{m^2_i(RGE)}{m^2_i}
\right) \quad \forall i 
\eeq
From \Eq{eta_def} we can easily see that $\eta \ge 1$. 
We are then all set for a minimization procedure. We were not able to
find solutions with strict universality. But if we 
relaxed\footnote{This meant that the $C's$ in \Eq{universality} were
not equal to 1. A few percent of non--universality was enough to get 
solutions.} the universality conditions on the squared masses of the 
singlet fields we got plenty of solutions.

\subsection{Main Features of the Model}

In this section we will review the main features of the model of
spontaneous broken $R_p$ described by Eqs.~(\ref{W_RP}) and (\ref{V_SOFT}).

\subsubsection{Chargino Mass Matrix}

The form of the chargino mass matrix is common to 
a wide class of  SUSY models with spontaneously broken $R_p$
and is given by~\cite{Nogueira90,RPSUSY}

\beq
\begin{array}{c|cccccccc}
& e^+_j & \tilde{H^+_u} & -i \tilde{W^+}\cr
\hline
\vb{20}
e_i & h_{e ij} v_d & - h_{\nu ij} v_{Rj} &  g v_{Li} \cr
\vb{20}
\tilde{H^-_d} & - h_{e ij} v_{Li} & \mu & g v_d \cr
\vb{20}
-i \tilde{W^-} & 0 &  g v_u & M_2 
\end{array}
\eeq

Two matrices U and V are needed to diagonalize the $5 \times 5$ 
(non-symmetric) chargino mass matrix
\beqa
{\chi}_i^+ &=& V_{ij} {\psi}_j^+ \qquad ; \qquad
\psi_j^+ = (e_1^+, e_2^+ , e_3^+ ,\tilde{H^+_u}, -i \tilde{W^+})\cr
\vb{18}
{\chi}_i^- &=& U_{ij} {\psi}_j^- \qquad ; \qquad 
\psi_j^- = (e_1^-, e_2^- , e_3^-, \tilde{H^-_d}, -i\tilde{W^-})
\label{INO}
\eeqa
where the indices $i$ and $j$ run from $1$ to $5$.

\subsubsection{Neutralino Mass Matrix}

Under reasonable approximations, we can truncate the neutralino 
mass matrix so as to obtain an effective $7\times 7$ matrix~\cite{RPSUSY}

\beq
\begin{array}{c|cccccccc}
& {\nu}_i & \tilde{H}_u & \tilde{H}_d & -i \tilde{W}_3 & -i \tilde{B}
\nonumber \\
\hline
{\nu}_i & 0 & h_{\nu ij} v_{Rj} & 0 & \frac{\ds g}{\ds \sqrt{2}}
v_{Li} & -\frac{\ds g'}{\ds \sqrt{2}} v_{Li}\nonumber \\
\vb{18}
\tilde{H}_u & h_{\nu ij} v_{Rj} & 0 & - \mu 
& -\frac{\ds g}{\ds \sqrt{2}} v_u & \frac{\ds g'}{\ds \sqrt{2}} v_u\nonumber \\
\vb{18}
\tilde{H}_d & 0 & - \mu & 0 & \frac{\ds g}{\ds \sqrt{2}} v_d 
& -\frac{\ds g'}{\ds \sqrt{2}} v_d\nonumber \\
\vb{18}
-i \tilde{W}_3 & \frac{\ds g}{\ds \sqrt{2}} v_{Li} 
& -\frac{\ds g}{\ds \sqrt{2}} v_u & \frac{\ds g}{\ds \sqrt{2}} v_d 
& M_2 & 0\nonumber \\
\vb{18}
-i \tilde{B} & -\frac{\ds g'}{\ds \sqrt{2}} v_{Li} & \frac{\ds g'}{\ds
\sqrt{2}} v_u & -\frac{\ds g'}{\ds \sqrt{2}} v_d & 0 & M_1 \nonumber
\end{array}
\label{nino}
\eeq

This matrix is diagonalized by a $7 \times 7$ unitary matrix N,
\beq
{\chi}_i^0 = N_{ij} {\psi}_j^0 \qquad \hbox{where} \qquad
\psi_j^0 = ({\nu}_i,\tilde{H}_u,\tilde{H}_d,-i \tilde{W}_3,-i
\tilde{B})
\eeq

\subsubsection{Charged Current Couplings}

Using the diagonalization matrices we can write the 
charged current Lagrangian describing the 
weak interaction between charged lepton/chargino and
neutrino/neutralinos  as
\beq
\lag^{CC}=\frac{g}{\sqrt2} W_\mu \bar{\chi}_i^- \gamma^\mu 
(K_{Lik} P_L + K_{Rik} P_R) {\chi}_k^0 + h.c.
\label{CC}
\eeq
where the $5\times 7$ coupling matrices $K_{L,R}$ may
be written as

\beqa
K_{Lik} &=& \eta_i (-\sqrt2 U_{i5} N_{k6} - U_{i4}
N_{k5} - \sum_{m=1}^{3} U_{im}N_{km})\label{KL}\cr
\vb{18}
K_{Rik} &=& \epsilon_k (-\sqrt2 V_{i5} N_{k6} + V_{i4} N_{k4})
\eeqa

\subsubsection{Neutral Current Couplings}

The corresponding neutral current Lagrangian may be written as
\beq
\label{NC}
\lag^{NC}=\frac{g}{\cos\theta_W}\ Z_\mu \left[ \bar{\chi}_i^- \gamma^\mu 
(O'_{Lik} P_L + O'_{Rik} P_R) \chi_k^- 
+ \frac{1}{2} \
\bar{\chi}_i^0 \gamma^\mu 
( O''_{Lik} P_L + O''_{Rik} P_R) \chi_k^0 
\right]
\eeq
where the $7 \times7$ coupling matrices $O'_{L,R}$ and
$O''_{L,R}$ are given by
\beqa
O'_{Lik}
&\hskip -3mm=\hskip -3mm&
\eta_i \eta_k \hskip -2mm
\left( \frac{1}{2} U_{i4} U_{k4} + U_{i5}
U_{k5} + \frac{1}{2} \sum_{m=1}^{3} U_{im}U_{km} -\hskip -1mm \delta_{ik}
\sin^2\theta_W \right)\label{OL}\cr
\vb{18}
O'_{Rik} 
&\hskip -3mm=\hskip -3mm&
\frac{1}{2} V_{i4} V_{k4} + V_{i5}
V_{k5} - \delta_{ik} \sin^2\theta_W \label{OR}\cr
\vb{18}
O''_{Lik} 
&\hskip -3mm=\hskip -3mm&
\frac{1}{2} \epsilon_i \epsilon_k   \left( N_{i4} N_{k4} - N_{i5}
N_{k5} - \sum_{m=1}^{3} N_{im}N_{km} \right)\hskip -2mm =
\hskip -2mm - \epsilon_i \epsilon_k
O''_{Rik} 
\eeqa
In writing these  couplings we have assumed CP conservation. 
Under this assumption the diagonalization matrices can be chosen 
to be real. The $\eta_i$ and $\epsilon_k$ factors are sign 
factors, related with the relative CP parities of these 
fermions, that follow from the diagonalization of their 
mass matrices.

\subsubsection{Parameters values}

All the results discussed in the following sections use Eqs.~(\ref{CC})
and (\ref{NC}) for the charged and neutral currents, respectively. To
compare with the experiment we need to discuss the input parameters.
Typical values for the SUSY parameters $\mu\equiv h_0 \vev{\Phi}$, $M_2$, 
and the parameters $h_{\nu i,3}$ lie in the range  
\beq
\begin{array}{lccl}
-250\leq\frac{\displaystyle\mu}{\mbox{GeV}}\leq 250 & & & 
30 \leq \frac{\displaystyle M_2}{\mbox{GeV}}\leq 1000 \cr
\vb{20}
10^{-10}\leq h_{\nu 13},h_{\nu 23} \leq 10^{-1} & & & 
10^{-5}\leq h_{\nu 33} \leq 10^{-1}\nonumber \\
\end{array}
\eeq
and we take the GUT relation $M_1/M_2=5/3\, 
\tan^2 \theta_W$.
For the expectation values we take the following range:

\beq
\begin{array}{lccl}
v_L=v_{L3}=100\:\: \mbox{MeV} & & & 
v_{L1}=v_{L2}=0 \cr
\vb{20}
50\:\: \mbox{GeV}\leq v_R=v_{R3}\leq 1000 \:\: \mbox{GeV} \ & & & 
v_{R1}=v_{R2}=0\cr
\vb{20}
50\:\: \mbox{GeV}\leq v_S=v_{S3}=v_R\leq 1000 \:\: \mbox{GeV} 
& & &   v_{S1}=v_{S2}=0\cr
\vb{20}
1 \leq \tan\beta=\ds  \frac{v_u}{v_d} \leq 50  
\end{array}
\eeq
which means that in practice we are considering that $R_p$ breaking is
obtained only through $\tau$ lepton number violation.

\subsubsection{Experimental Constraints}

Before we close this section on the spontaneously broken $R_p$ model
we have to discuss what are the experimental constraints on the
model. Some of these  constraints are  common to all SUSY models, and
are related to the negative results of the searches for the
superpartners. This in turn puts constraints in the parameters of the
models. But there are other constraints that are more characteristic
of the spontaneously broken $R_p$ models, in particular those that
are related to lepton flavor violation. We will give here a short
list of the constraints that we have been using.

\begin{itemize}

\item {\bf LEP searches}

The most recent limits on 
chargino masses from the recent runs
were included.

\item {\bf Hadron Colliders}

From $\bar pp$ colliders there are restrictions on gluino 
production and hence on the gluino mass.

\item {\bf Non--Accelerator Experiments}

They follow from laboratory experiments related to {\sl neutrino physics}, 
{\sl cosmology} and {\sl astrophysics}.
The most relevant are:

\begin{itemize}

\item 
Neutrinoless double beta decay 

\item
Neutrino oscillation searches 

\item
Direct searches for anomalous peaks at $\pi$
and K meson decays

\item
The limit on the tau neutrino mass

\item
Cosmological limits on the $\nu_{\tau}$ lifetime and mass 

\end{itemize}

\end{itemize}

\subsection{Implications for Neutrino Physics}

Here we briefly summarize the main results for neutrino physics.

\begin{itemize}

\item
{\it Neutrinos have mass}

Neutrinos are massless at Lagrangian level but get mass from the
mixing with neutralinos\cite{Nogueira90,RPSUSY}.

\item
{\it Neutrinos mix}

The coupling matrix $h_{\nu_{ij}}$ has to be non diagonal to allow
\beq
\nu_{\tau} \ra \nu_{\mu} + J 
\eeq
and therefore evading~\cite{RPSUSY} the {\it Critical Density Argument} against
$\nu's$ in the MeV range. The fact that $h_{\nu_{ij}}$ has to be non
diagonal leads to important consequences in lepton violating processes
as we will see below.

\item
{\it Avoiding BBN constraints on the $m_{\nu_{\tau}}$}

In the SM BBN arguments~\cite{bbnothers} rule out $\nu_{\tau}$ 
masses in the range
\beq
0.5\ MeV < m_{\nu_{\tau}} < 35\ MeV
\eeq
We have shown~\cite{bbnpaper} that SBRP models can evade that 
constraint due to new annihilation channels
\beq
\nu_{\tau} \nu_{\tau} \ra J J 
\eeq

\end{itemize}

\subsection{R--parity in Non--Accelerator Experiments}

Here we will describe the implications of SBRP in non accelerator
experiments like the solar neutrinos experiments and flavor violating
leptonic decays.

\subsubsection{Solar Neutrinos}

To a good approximation we can write~\cite{RPSUSY}
\begin{eqnarray}
\nu_1&=&\cos \theta \nu_e -\sin \theta \nu_{\mu}\cr
\vb{14}
\nu_2&=&\sin \theta \nu_e +\sin \theta \nu_{\mu} \cr
\vb{14}
\nu_3&=&\nu_{\tau} 
\end{eqnarray}
where $\nu_i$, $i=1,2,3$ and $\nu_e$, $\nu_{\mu}$ and $\nu_{\tau}$ are,
respectively, the mass and weak interaction eigenstates. The mixing angle
$\theta$ is given in terms of the model parameters by
\beq
\tan \theta ={h_{\nu_{13}} \over h_{\nu_{23}}}
\eeq
The constraints on $h_{\nu_{13}}$ and $h_{\nu_{23}}$
do not restrict much their ratio. Therefore a large range of mixing angles 
is allowed. For the masses we get~\cite{RPSUSY}
\beqa
&m_1&=0 \cr
\vb{15}
10^{-4} eV \le &m_2& \le 10^{-2} eV \cr
\vb{15}
10 keV \le &m_{\tau}& \le 23 MeV
\eeqa
just in the right range for the MSW mechanism.

\subsubsection{SUSY Signals in $\mu$ and $\tau$ Decays}

The existence of a massless scalar particle, the majoron,
can affect the decay spectra of the $\mu$ and $\tau$ leptons
through the emission of the Majoron in processes such as

\beq
\mu\ra e +  J \qquad ; \qquad
\tau\ra e +  J \qquad ; \qquad
\tau\ra \mu +  J 
\eeq
These are flavor violating decays that are present in our model because
the matrix $h_{\nu i j}$ is not flavor diagonal. After a careful
sampling of the parameter space, we found out that the rates can be
close to the present experimental limits~\cite{Nuria}. For instance
for the process $\mu \ra e J$ we can go up to the present experimental
limit~\cite{muej}, $BR(\mu \ra e J)< 2.6 \times 10^{-6}$.

\subsection{R--parity Violation at LEP I}

\subsubsection{Higgs Physics}

The structure of the neutral Higgs sector is more complicated then in
the MSSM. However the main points are simple.

\begin{itemize}

\item
{\it Reduced Production}

Like in the MSSM the coupling of the Higgs to the $Z^0$ is
reduced by a factor $\epsilon_B$
\beq
\epsilon_B=\left\vert 
\frac{g_{ZZh}}{\vbox to 12pt {} g_{ZZh}^{SM}} \right\vert < 1
\eeq

\item
{\it Invisible decay}

Unlike the SM and the MSSM where the Higgs decays mostly in 
$b\overline{b}$, here it can have {\it invisible} decay modes like
\beq
H \ra J + J 
\eeq
Depending on the parameters, the $BR(H\ra \hbox{invisible})$ can be large.
This will relax the mass limits obtained from LEP. 
We performed a model
independent analysis of the LEP data~\cite{higgs} 
taking $m_H$, $\varepsilon_B$
and $BR(H\ra \hbox{invisible})$ as independent parameters. The results
are shown in Fig.~(\ref{val95_f4}a)

\begin{figure}
\begin{center}
\begin{tabular}{cc}
\vspace{0cm}
\includegraphics[height=6cm]{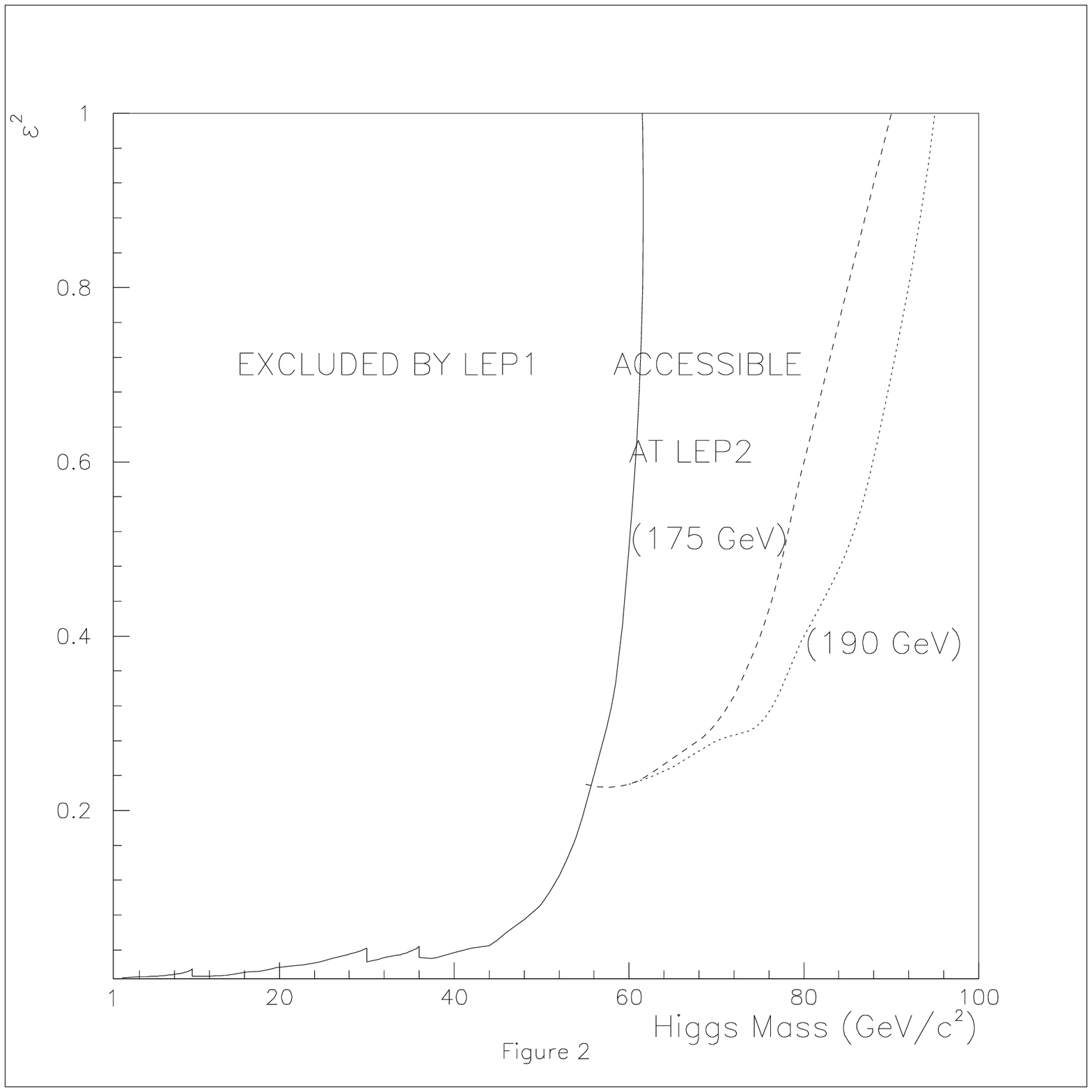}
&
\vspace{0cm}
\includegraphics[bb=26 200 533 720,height=6.5cm]{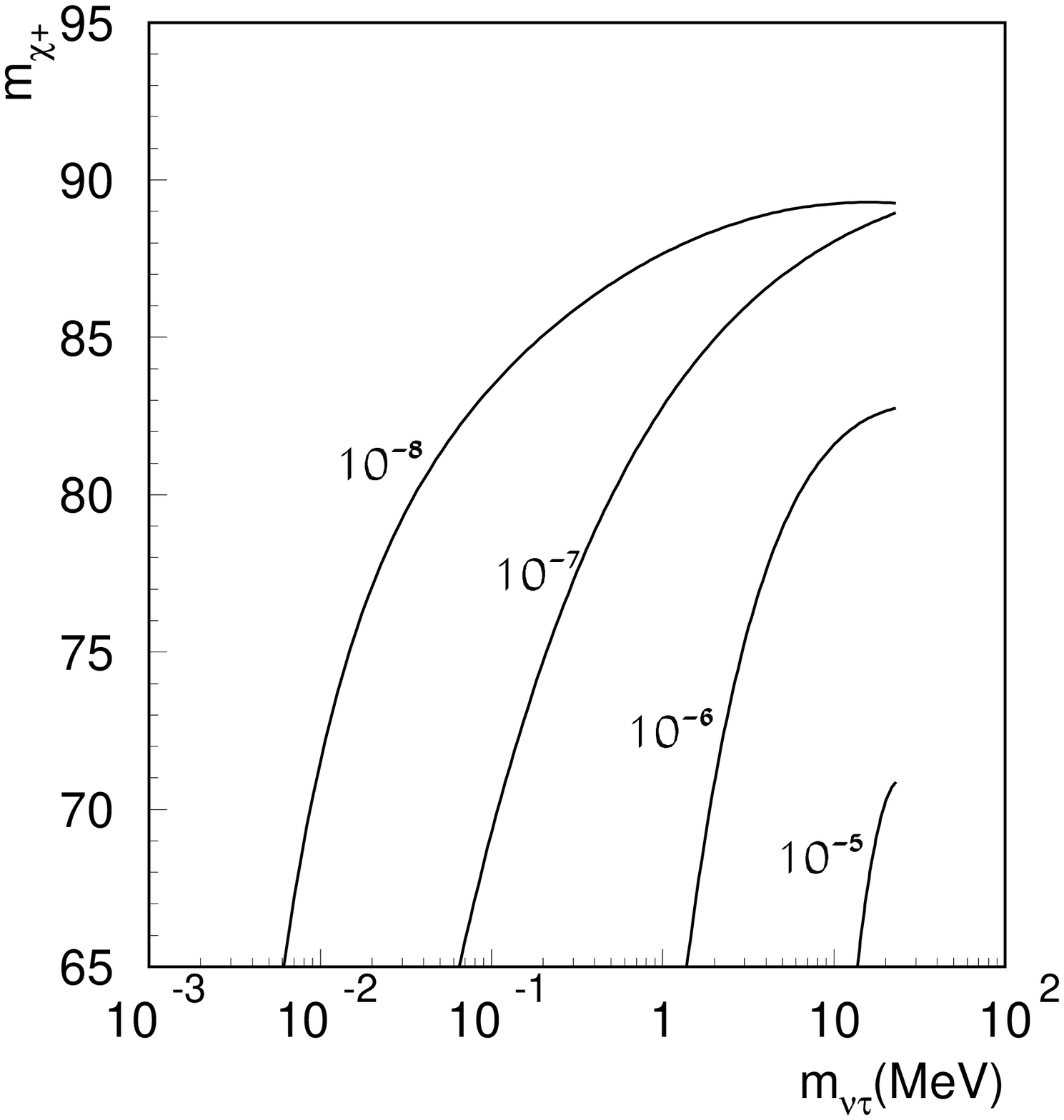}\cr
a)&b)
\end{tabular}
\end{center}
\vspace{-5mm}
\caption{\small
a) Limits on the $\epsilon^2$ versus $m_H$ plane obtained from LEP, 
b) Attainable $BR(Z^0\ra \chi^{\pm} \tau^{\mp}$) as a
function of the chargino and $\nu_{\tau}$ masses.}
\label{val95_f4}
\end{figure}

\end{itemize}

\subsubsection{Chargino Production at the Z Peak}
\label{sbrpchitau}

The more important is the possibility of the decay 
\beq
Z^0 \to \chi^{\pm} \tau^{\mp}
\eeq
This decay is possible because $R_p$ is broken. We have 
shown~\cite{Nogueira90,val95} that this branching ratio can be as high 
as $5\times 10^{-5}$. This is shown in Fig.~(\ref{val95_f4}b). 
Another important point is that the chargino has different decay modes
with respect to the MSSM. 
\beqa
\chi &\ra& \chi^0 + f \overline{f'}\cr
\vb{20}
\chi &\ra&\tau + J
\eeqa
The relative importance of the 2--body over the 3--body is very much
dependent on the parameters of the model, but the 2--body can dominate.

\subsubsection{Neutralino Production at the Z Peak}

We have developed an event generator that simulates the 
processes expected for the LEP collider at $\sqrt{s}=M_{Z}$. Its main
features are:

\begin{itemize}

\item
{\it Production}

As far as the production is concerned, our generator 
simulates the following processes at the $Z$ peak:
\bea
&&\ e^+ e^-\rightarrow \chi\nu \\
&&\ e^+ e^-\rightarrow \chi \chi
\eea
\item
{\it Decay}

The second step of the generation is the decay of the lightest
neutralino. The 2-body only contributes to the missing energy. The
3-body are:
\bea
&&\chi \rightarrow \nu_{\tau} Z^{*} \rightarrow \nu_{\tau}\ l^+l^- ,
\nu_{\tau} \nu \nu,\nu_{\tau} q_i \overline{q_i}\\
&&\chi \rightarrow \tau \ W^{*} \rightarrow  \tau \nu_i l_i,\tau q_u
\overline{q_d}
\eea

\item
{\it Hadronization}

The last step of our simulation is made calling the 
PYTHIA software for the final states with quarks.

\end{itemize}
One of the cleanest and most interesting signals that can be 
studied is the process with missing transverse momentum + 
acoplanar muons pairs~\cite{RPVLEPI}
\beq
 p\!\!\!/_T +\mu^+ \mu^- 
\eeq
The main source of 
background for this signal is the 
\beq
Z \ra \mu^+\mu^- + \hbox{soft photons}
\eeq
For definiteness we have imposed the cuts used by the OPAL 
experiment for their search for acoplanar dilepton events:
(a) We select events with two muons with at least for one of the muons
obeying $ |\cos\theta | $ less than 0.7. 
(b) The energy of each muon has to be greater than a $6\%$ of the beam
energy.
(c) The missing transverse momentum in the event must exceed
$6\%$ of the beam energy, $p\!\!\!/_T > 3 \:$ GeV.
(d) The acoplanarity angle (the angle between the projected
momenta of the two muons in the plane orthogonal to the 
beam direction) must exceed $20^o$. With these cuts we were able to
calculate the efficiencies of our processes.

\begin{figure}
\begin{center}
\begin{tabular}{cc}
\includegraphics[height=75mm]{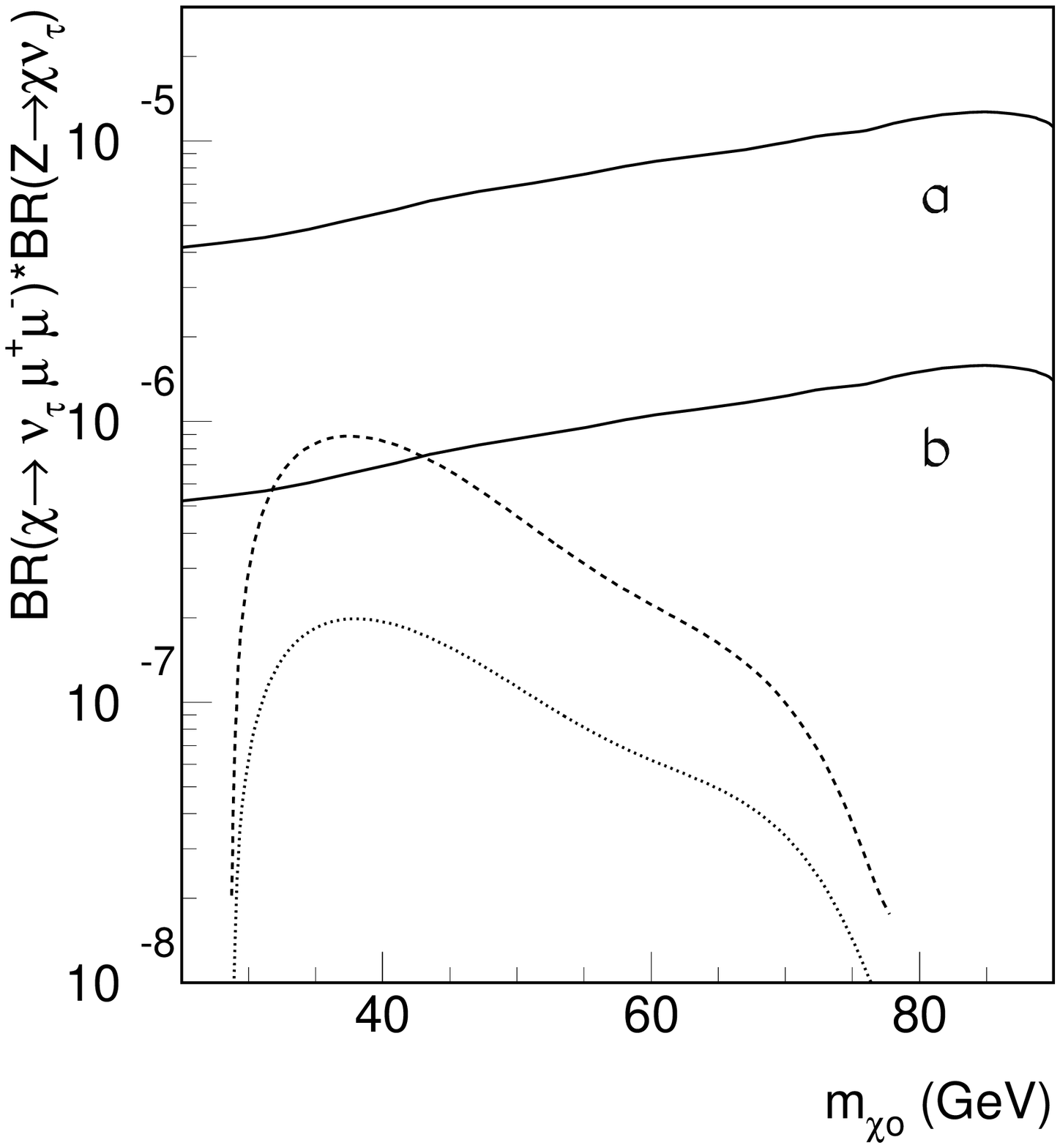}
&
\includegraphics[height=75mm]{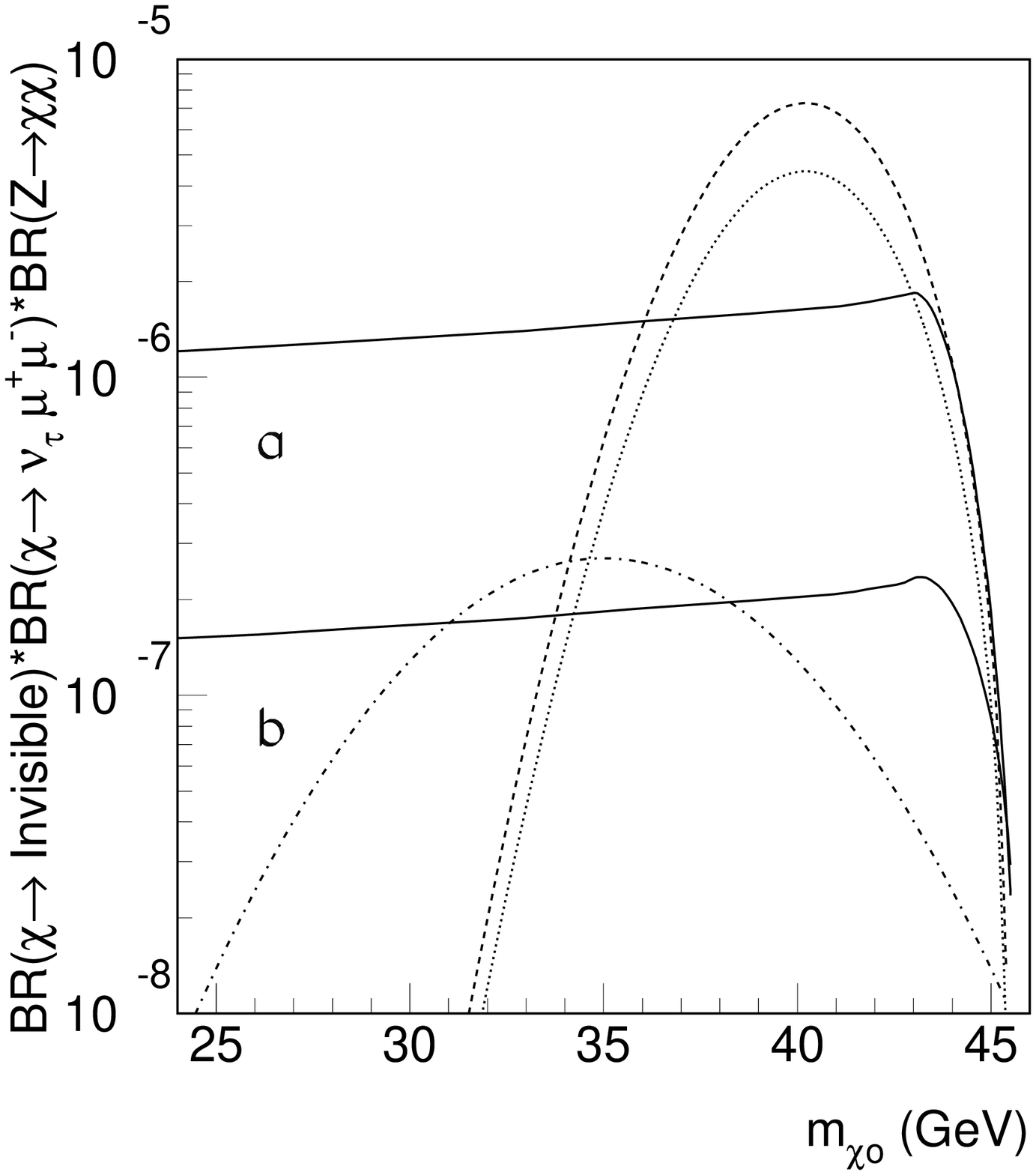}
\end{tabular}
\end{center}
\vspace{-8mm}
\caption{\small
On the left a comparison of the attainable limits on
$BR (Z \ra \chi \nu) BR(\chi \rightarrow \mu^+\mu^- \nu)$
versus the lightest neutralino mass, with the maximum theoretical values 
expected in different R--parity breaking models. The solid line (a) is
just for the $\mu^+\mu^-\nu$ channel, 
while (b) corresponds to the improvement expected
from including the $e^+e^-\nu$ channel, as well as the 
combined statistics of the four LEP experiments.
The dashed line corresponds to a model with explicit R--parity
violation,while the dotted one is calculated in the 
spontaneous R--parity-violation model. On the right the same for
$BR (Z \ra \chi \chi) BR(\chi \rightarrow \mu^+\mu^- \nu)$.
}
\label{fig1}
\end{figure}

\noindent
We used the data published by ALEPH in 95 and analyzed both the single
production $e^+e^- \rightarrow \chi \nu $ and the double production 
$e^+e^- \rightarrow \chi \chi$
processes. For single production we get
\beq
N_{expt} (\chi \nu)  = \sigma(e^+e^-\rightarrow \chi \nu) BR(\chi \rightarrow 
\nu_\tau \mu^+\mu^-) \epsilon_{\chi \nu} \  L_{int}
\eeq
Using the expression for the cross section we can write this
expression in terms of the product $BR (Z \ra \chi \nu)$ $\times$
$BR(\chi \rightarrow \nu_\tau \mu^+ \mu^-)$
and obtain a 
$95\% CL $ limit on this R--parity breaking observable,
as a function of the $\chi$ mass. This is shown in 
Figure~\ref{fig1}. For the double production of neutralinos the number
of expected $p\!\!\!/_T+\mu^+ \mu^-$  events is
\beq
N_{expt} (\chi \chi) = \sigma (e^+e^- \rightarrow \chi \chi)
2 BR (\chi \rightarrow \mbox{invisible})
BR(\chi\rightarrow\nu_\tau\mu^+\mu^-) \epsilon_{\chi \chi} \  L_{int}
\eeq
We can obtain 
an illustrative $95\% CL$ limit on $BR (Z \ra \chi \chi)$ $\times$
$BR (\chi \ra \nu_\tau \mu^+\mu^-)$ $\times$ $ BR (\chi \ra \mbox{invisible})$ 
as a function of the $\chi$ mass~\cite{RPVLEPI}. This is also shown in Figure \ref{fig1}
where we can see that the models begin to be constrained by the LEP results.

\subsection{R--parity Violation at LEP II}

\subsubsection{Invisible Higgs}

The previous LEP I analysis has been extended for LEP II.\cite{eboli}
As a general framework we consider models with the  interactions
\begin{eqnarray}
{\cal L}_{hZZ}
&=& \epsilon_B \left ( \sqrt{2}~ G_F \right )^{1/2}
M_Z^2 Z_\mu Z^\mu h 
\; , \cr
\vb{18}
{\cal L}_{hAZ}&=& - \epsilon_A \frac{g}{\cos\theta_W} 
Z^\mu h \stackrel{\leftrightarrow}{\partial_\mu} A 
\; ,
\end{eqnarray}
with $\epsilon_{A(B)}$ being determined once a model is chosen. We
also consider the possibility that the Higgs decays invisible
\beq
h \ra JJ
\eeq
and treat the branching fraction $B$ for $h \rightarrow JJ$ as a 
free parameter.  

\begin{figure}[ht]
\begin{center}
\begin{tabular}{cc}
\includegraphics[width=70mm,bbllx=10pt,bblly=10pt,bburx=530pt,bbury=530pt]{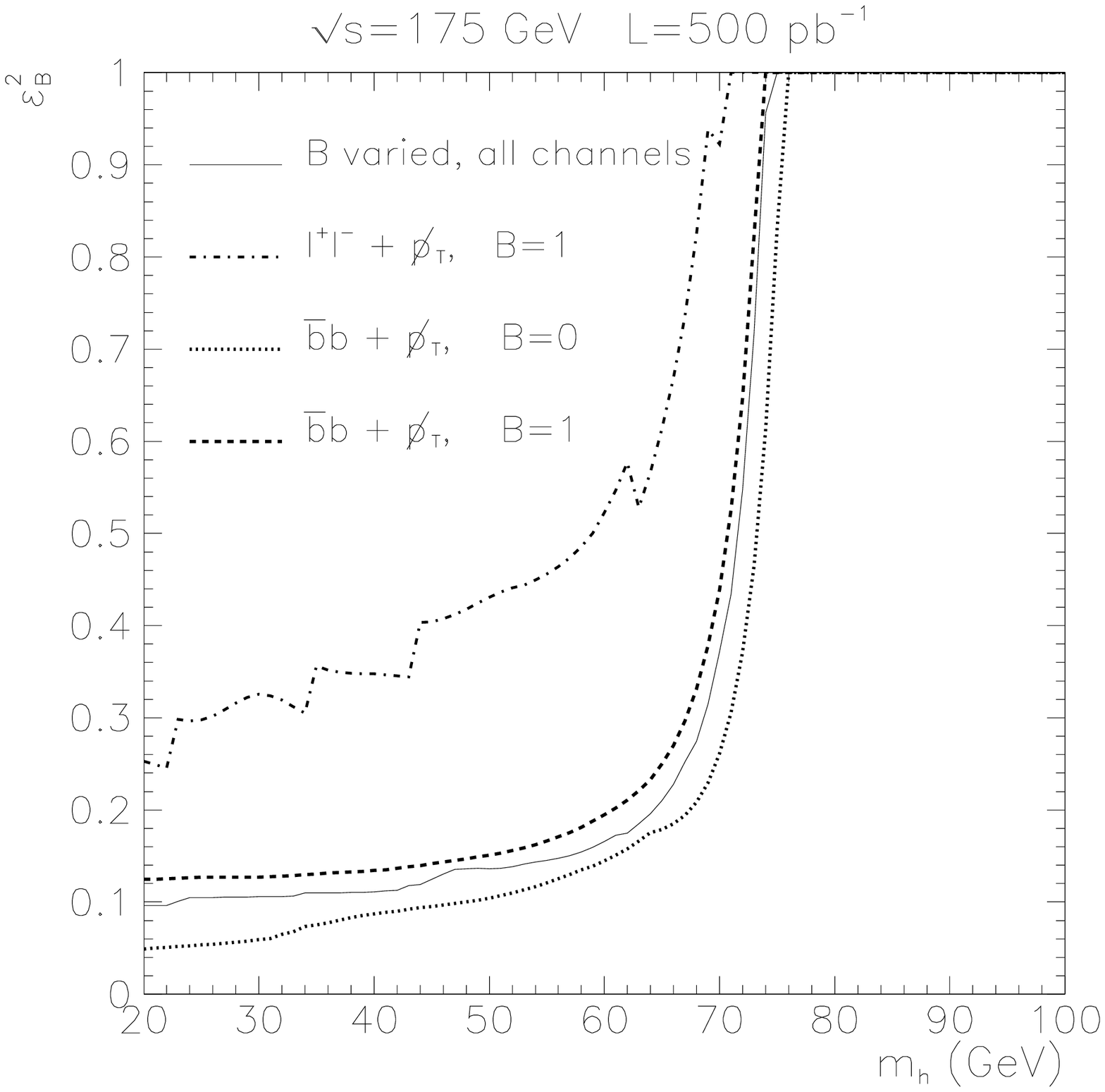}
&
\includegraphics[width=70mm,bbllx=10pt,bblly=10pt,bburx=530pt,bbury=530pt]{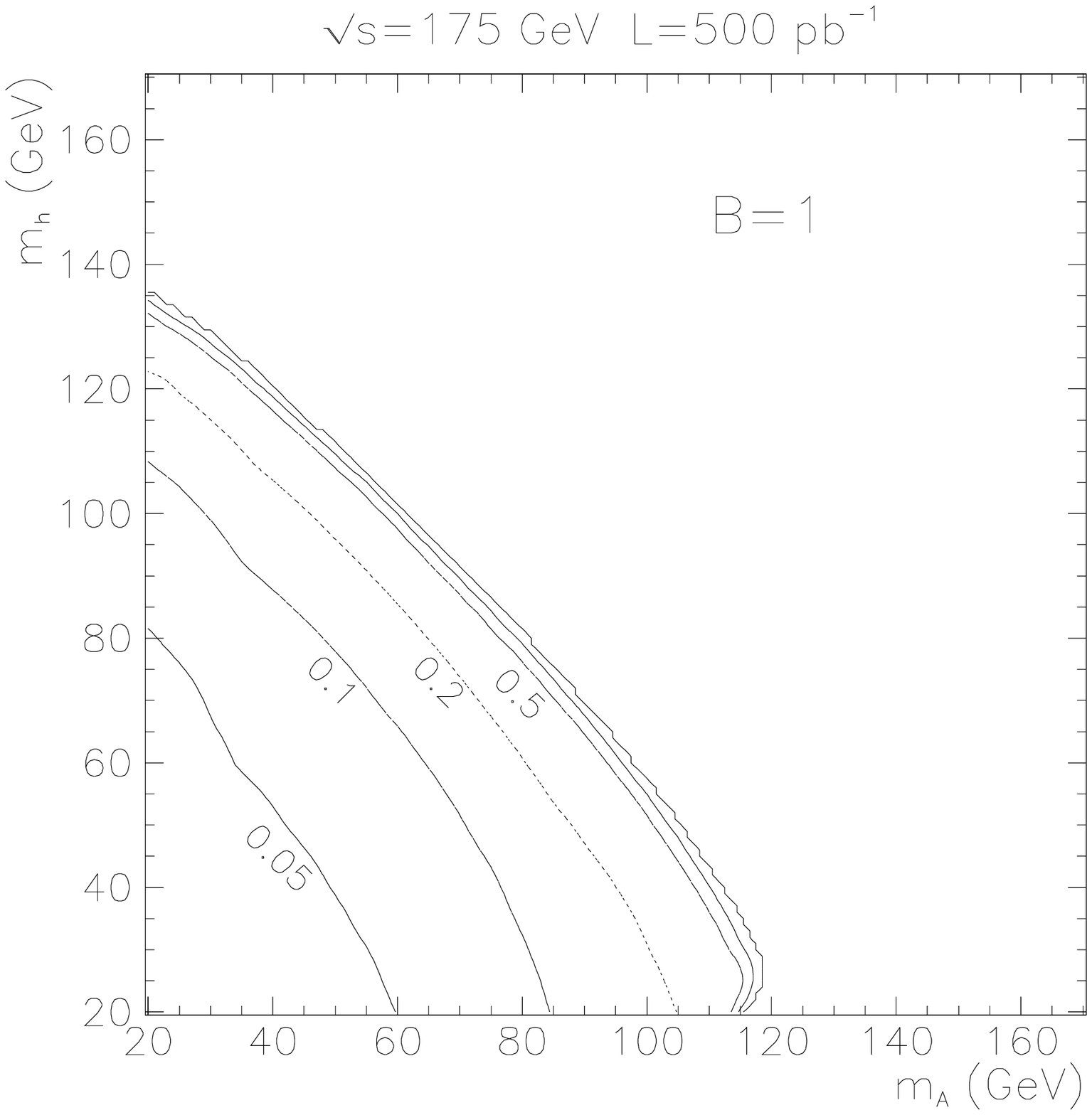}
\end{tabular}
\end{center}
\vspace{-5mm}
\caption{\small
On the left, bounds on $\epsilon_B^2$ as a function of $M_h$
for $\sqrt{s}=175 GeV$. On the right, bounds on $\epsilon_A^2$ as a
function of $M_h$ and $ M_A$ for $B=1$ and $\surd{s}= 175$ GeV.}
\label{fig3}
\end{figure}

\noindent
The following signals with $\ptmis$ were considered:
\begin{eqnarray}
e^+ e^-  &\rightarrow & (Z h + A h) \rightarrow b \bar{b}~+~ \ptmis
\; ,\cr
\vb{18}
e^+ e^-  &\rightarrow & Z h \rightarrow \ell^+ \ell^- ~+~ \ptmis
\; ,
\end{eqnarray}
but also the more standard processes
\begin{eqnarray}
e^+ e^-  &\rightarrow & Z h \rightarrow \ell^+ \ell^-  +~b \bar{b} 
\; ,\cr
\vb{18}
e^+ e^-  &\rightarrow & (Z h + A h) \rightarrow b \bar{b}~ +~b \bar{b}
\; .
\end{eqnarray}
Using the above processes and after a careful study of the backgrounds
and of the necessary cuts,~\cite{eboli} it was possible to 
evaluate the limits on $M_h$, $M_A$, $\epsilon_A$,
$\epsilon_B$, and $B$ that can be obtained at LEP II. In
Figure~\ref{fig3}  are shown some of these limits.

\subsubsection{Neutralinos and Charginos}

At LEP II the production rates for {\it R--parity violation} processes
will not be very large, compared with
those at LEP I. Therefore we expect that the production rates will be like
in the MSSM, via non R--parity breaking processes. 
However the decays will be
modified much in the same way as in the LEP I case. This is specially
important for the $\chi_0$ because it is invisible in the MSSM but visible
here. 
Also the R--parity violating decays of the charginos
\beq
\chi^- \ra \tau^- + J
\eeq
can have a substantial decay fraction compared with the usual MSSM decays
\beq
\chi^- \ra \chi^0 + f \overline{f'}
\eeq

\newpage

\section{Bilinear R--parity Violation: The $\epsilon$ model}

We have seen in the previous section that it could well be that R--parity is a 
symmetry at the Lagrangian level but is broken by the ground state.  
Such scenarios provide a very {\sl 
systematic} way to include R parity violating effects, automatically 
consistent with low energy {\sl baryon number conservation}. They have 
many added virtues, such as the possibility of providing a dynamical 
origin for the breaking of R--parity, through radiative corrections, 
similar to the electroweak symmetry \cite{rprad}.  The simplest 
truncated version of such a model, in which the violation of R--parity 
is effectively parameterized by a bilinear superpotential term 
$\epsilon_i\widehat L_i^a\widehat H_2^b$ has been widely discussed 
\cite{RPeps,epsrad}. It has also been shown recently \cite{epsrad} that this 
model is consistent with minimal N=1 supergravity unification with 
radiative breaking of the electroweak symmetry and universal scalar 
and gaugino masses. This one-parameter extension of the MSSM-SUGRA 
model therefore provides the simplest reference model for the breaking 
of R--parity and constitutes a consistent truncation of the complete 
dynamical models with spontaneous R--parity breaking proposed 
previously \cite{Masiero90}.  In this case there is no physical Goldstone 
boson, the Majoron, associated to the spontaneous breaking of 
R--parity, since in this effective truncated model the superfield 
content is exactly the standard one of the MSSM. Formulated as an 
effective theory at the weak scale, the model contains only two new 
parameters in addition to those of the MSSM. Therefore our model 
provides also the simplest parameterization of R--parity breaking 
effects.  In contrast to models with tri-linear R--parity breaking 
couplings, it leads to a very restrictive and systematic pattern of 
R--parity violating interactions, which can be taken as a reference 
model. 
In this section we will review the most important features of this
model.

\subsection{Description of the Model}

The superpotential $W$ is given by 
\begin{eqnarray}
W&=&\varepsilon_{ab}\left[
 h_U^{ij}\widehat Q_i^a\widehat U_j\widehat H_2^b
+h_D^{ij}\widehat Q_i^b\widehat D_j\widehat H_1^a
+h_E^{ij}\widehat L_i^b\widehat R_j\widehat H_1^a 
-\mu\widehat H_1^a\widehat H_2^b
+\epsilon_i\widehat L_i^a\widehat H_2^b\right]
\end{eqnarray}
where $i,j=1,2,3$ are generation indices, $a,b=1,2$ are $SU(2)$
indices. In the following we will consider, for simplicity, only the third
generation. Then the set of soft supersymmetry
breaking terms are
\begin{eqnarray}
V_{soft}&\hskip -3mm=\hskip -3mm&
M_Q^{2}\widetilde Q^{a*}_3\widetilde Q^a_3+M_U^{2}
\widetilde U^*_3\widetilde U_3+M_D^{2}\widetilde D^*_3
\widetilde D_3+M_L^{2}\widetilde L^{a*}_3\widetilde L^a_3+
M_R^{2}\widetilde R^*_3\widetilde R_3
+m_{H_1}^2 H^{a*}_1 H^a_1\cr
&&\vb{20}\hskip -5mm
+m_{H_2}^2 H^{a*}_2 H^a_2 
- \left[\half M_3\lambda_3\lambda_3+\half M_2\lambda_2\lambda_2
+\half M_1\lambda_1\lambda_1+h.c.\right]\cr
&&\vb{20}\hskip -5mm
+\varepsilon_{ab}\! \left[\!
A_th_t\widetilde Q^a_3\widetilde U_3 H_2^b
+\! A_bh_b\widetilde Q^b_3\widetilde D_3 H_1^a
+\! A_{\tau}h_{\tau}\widetilde L^b_3\widetilde R_3 H_1^a 
\! -\! B\mu H_1^a H_2^b+\! B_2\epsilon_3\widetilde L^a_3 H_2^b\right]
\end{eqnarray}
The bilinear
$R_p$ violating term {\sl cannot} be eliminated by superfield
redefinition.
The reason is that the bottom Yukawa coupling, usually neglected,
plays a crucial role in splitting
the soft-breaking parameters $B$ and $B_2$ as well as the scalar
masses $m_{H_1}^2$ and $M_L^{2}$, assumed to be equal at the
unification scale.

The electroweak symmetry is broken when the VEVS of 
the two Higgs doublets $H_1$
and $H_2$, and the tau--sneutrino.
\beq
H_1=\left(\begin{array}{c}
{\ds \chi^0_1+v_1+i\varphi^0_1\over{\ds \sqrt{2}}}\cr
\vb{20}
H^-_1
\end{array}
\right)
\ , \
H_2=\left(\begin{array}{c}
H^+_2\cr
\vb{20}
{\ds \chi^0_2+v_2+
i\varphi^0_2\over{\ds \sqrt{2}}}
\end{array}
\right)
\ , \
\widetilde L_3=\left(\begin{array}{c}
{\ds \tilde\nu^R_{\tau}+v_3+i\tilde\nu^I_{\tau}\over{\ds \sqrt{2}}}\cr
\vb{20}
\tilde\tau^-
\end{array}
\right)
\eeq

The gauge bosons $W$ and $Z$ acquire masses
$m_W^2=\quarter g^2v^2$, $ m_Z^2=\quarter(g^2+g'^2)v^2$, where
\beq
v^2\equiv v_1^2+v_2^2+v_3^2=(246 \; {\rm GeV})^2
\eeq
We introduce the
following notation in spherical coordinates:
\begin{eqnarray}
v_1&=&v\sin\theta\cos\beta\cr
\vb{14}
v_2&=&v\sin\theta\sin\beta\cr
\vb{14}
v_3&=&v\cos\theta
\end{eqnarray}
which preserves the MSSM definition $\tan\beta=v_2/v_1$. 
The angle $\theta$ equal to $\pi/2$ in the MSSM limit.

The full scalar potential may be written as
\beq
V_{total}  = \sum_i \left| { \partial W \over \partial z_i} \right|^2
	+ V_D + V_{soft} + V_{RC}
\eeq
where $z_i$ denotes any one of the scalar fields in the
theory, $V_D$ are the usual $D$-terms, $V_{soft}$ the SUSY soft
breaking terms, and $V_{RC}$ are the 
one-loop radiative corrections. 
In writing $V_{RC}$ we  use the diagrammatic method and find 
the minimization conditions by correcting to one--loop the tadpole
equations. 
This method has advantages with respect to the effective potential when
we calculate the one--loop corrected scalar masses.
The scalar potential contains linear terms
\beq
V_{linear}=t_1^0\chi^0_1+t_2^0\chi^0_2+t_3^0\tilde\nu^R_{\tau}\,,
\eeq
where
\begin{eqnarray}
t_1^0&=&(m_{H_1}^2+\mu^2)v_1-B\mu v_2-\mu\epsilon_3v_3
+\eighth(g^2+g'^2)v_1(v_1^2-v_2^2+v_3^2)\,,
\cr
\vb{18}
t_2^0&=&(m_{H_2}^2+\mu^2+\epsilon_3^2)v_2-B\mu v_1+
B_2\epsilon_3v_3
-\eighth(g^2+g'^2)v_2(v_1^2-v_2^2+v_3^2)\cr
\vb{18}
t_3^0&=&(m_{L_3}^2+\epsilon_3^2)v_3-\mu\epsilon_3v_1+
B_2\epsilon_3v_2
+\eighth(g^2+g'^2)v_3(v_1^2-v_2^2+v_3^2)\,.
\label{eq:tadpoles}
\end{eqnarray}
These $t_i^0, i=1,2,3$ are the tree level tadpoles, and are equal to 
zero at the minimum of the potential. 

\subsection{Main Features}

\subsubsection{Charginos and Neutralinos}

\begin{figure}[ht]
\begin{center}
\begin{tabular}{cc}
\includegraphics[bb= 80 75 525
675,width=65mm,height=80mm,angle=90]{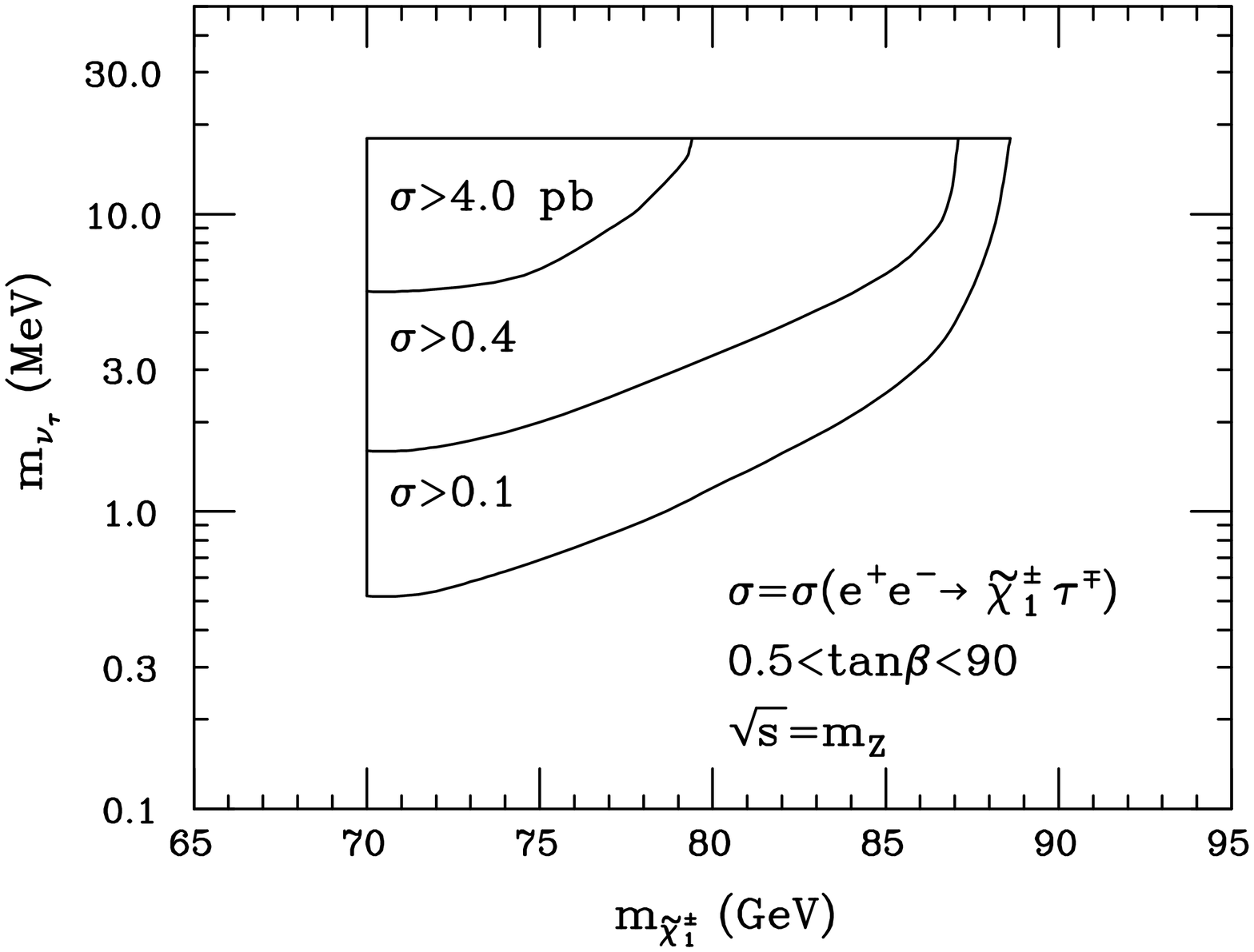}
&
\includegraphics[bb= 30 148 520 708,height=7cm]{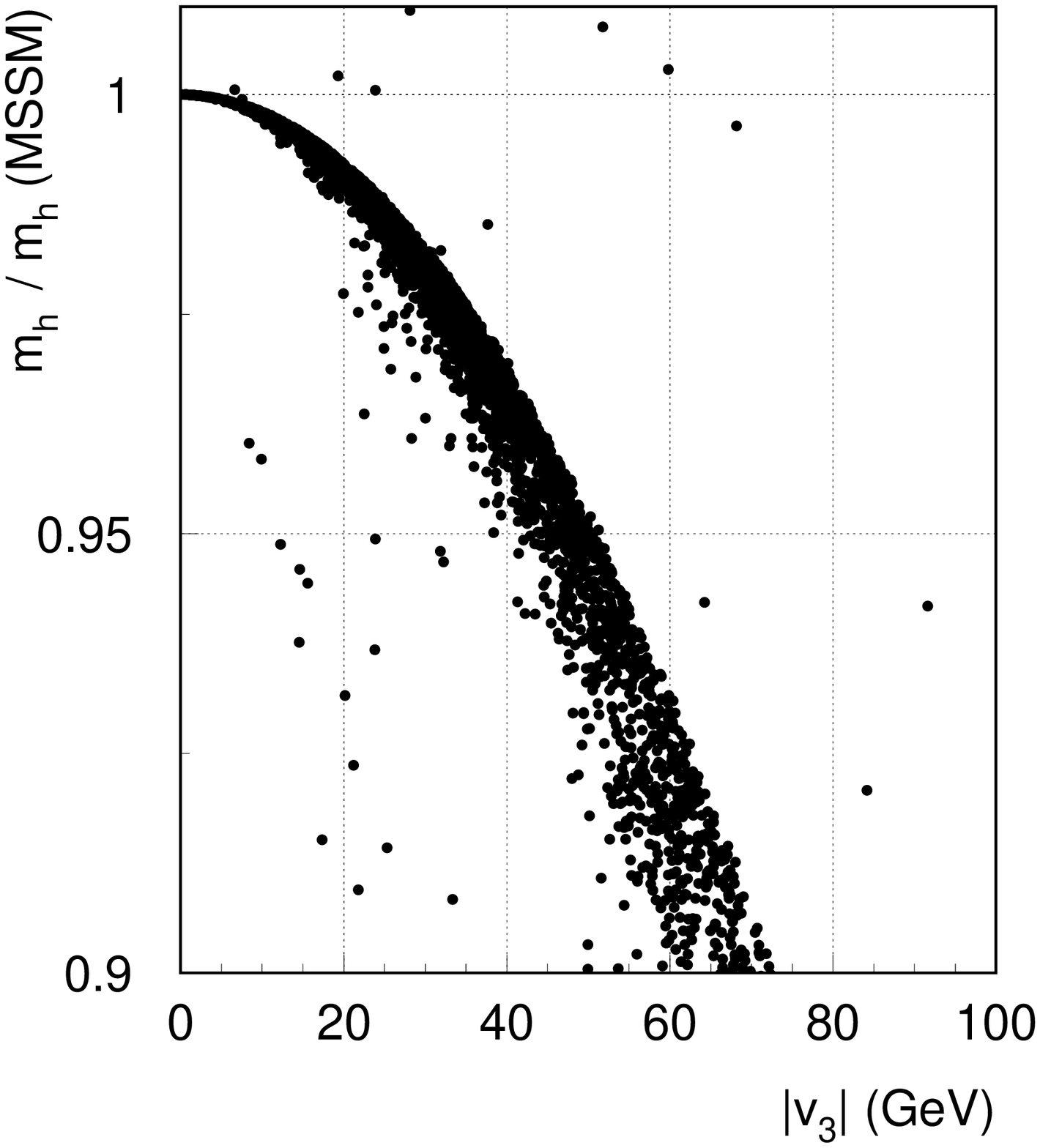}\cr
a)&b)
\end{tabular}
\end{center} 
\vspace{-5mm}
\caption{\small 
a) Regions of attainable cross section in BRpV in the plane 
tau neutrino mass {\it vs} chargino mass including large values of
$\tan\beta$. 
b)Ratio of the lightest CP--even Higgs boson mass in the 
$\epsilon$--model and in the MSSM  as a function of
$v_3$. 
}
\label{ratio} 
\end{figure}

The $\epsilon$--model is a one parameter generalization of the MSSM.
It can be thought as an {\sl effective} model
showing the more important features of the SBRP--model at the weak
scale. In fact the mass matrices, the charged and neutral currents, 
are similar to the SBRP--model if we identify
\beq
\epsilon_3 \equiv v_R h_{\nu}{}_{33}
\label{epsequiv}
\eeq
Therefore all that we said about the SBRP--model in Section 2 also
applies here. In particular the implications of the mixing of the
$\tau$ lepton with charginos have been studied in
ref.~\cite{epschitau}. Their results are shown in Fig.~\ref{ratio}a, 
and are similar to those of Section~\ref{sbrpchitau} 
if we use the identification
of Eq.~(\ref{epsequiv}). 
The only difference arises in processes where the Majoron
plays an important role, because it is absent here. This has been
studied in full detail in refs.~\cite{RPVLEPI,RPeps}.

The other important feature it is that this model has the
MSSM as a limit. This can be illustrated in Fig.~\ref{ratio}b, 
where we show the ratio of the lightest CP-even Higgs
boson mass $m_h$ in the
$\epsilon$--model and in the MSSM  as a function of
$v_3$. As $v_3\ra 0$ the ratio goes to one.

%\includegraphics[bb= 80 75 525
%675,height=75mm,angle=90]{ntch_90.ps}

\subsubsection{Charged Scalars}
\label{ChargedScalars}

The charged scalar sector is also similar to the SBRP--model,
because the extra superfields needed in that case are all
neutral. Therefore the charged scalars are the charged Higgs bosons,
sleptons and squarks. Because of the breaking of R--parity the charged
Higgs bosons are mixed with the charged
sleptons. If we consider only the third generation, the mixing will
be with the staus. Although this sector is similar in the
$\epsilon$--model and in the SBRP--model, the overall analysis is
simpler in $\epsilon$--model because it has fewer parameters.

The mass matrix of the charged scalar sector follows from the 
quadratic terms in the scalar potential 
\begin{equation} 
V_{quadratic}=[H_1^-,H_2^-,\tilde\tau_L^-,\tilde\tau_R^-] 
\bold{M_{S^{\pm}}^2}\left[\matrix{H_1^+ \cr H_2^+ \cr \tilde\tau_L^+ \cr 
\tilde\tau_R^+}\right] + \cdots
\label{eq:Vquadratic} 
\end{equation}
For convenience reasons we will divide this $4\times4$ matrix into 
$2\times2$ blocks in the following way: 
\begin{equation} 
\bold{M_{S^{\pm}}^2}=\left[\matrix{ 
{\bold M_{HH}^2} & {\bold M_{H\tilde\tau}^{2T}} \cr 
\vb{18}
{\bold M_{H\tilde\tau}^2} & {\bold M_{\tilde\tau\tilde\tau}^2} 
}\right] 
\label{eq:subdivM} 
\end{equation} 
where the charged Higgs block is 
\beq
\label{eq:subMHH} 
{\bold M_{HH}^2}\!=\!\! 
\left[\matrix{ 
B\mu{{v_2}\over{v_1}}\! +\! \quarter g^2(v_2^2\!-\!v_3^2)
\!+\!\mu\epsilon_3 
{{v_3}\over{v_1}}\!+\!\half h_{\tau}^2v_3^2\!+\!{{t_1}\over{v_1}} 
& B\mu+\quarter g^2v_1v_2 \cr
\vb{20} 
B\mu+\quarter g^2v_1v_2 
&\!\! B\mu{{v_1}\over{v_2}}\!+\!\quarter g^2(v_1^2\!+\!v_3^2)
\!-\!B_2\epsilon_3 
{{v_3}\over{v_2}}\!+\!{{t_2}\over{v_2}} 
}\right] 
\eeq
and $h_{\tau}$ is the tau Yukawa coupling. This matrix  
reduces to the usual charged Higgs mass matrix in the MSSM when we  
set $v_3=\epsilon_3=0$ and we call $m_{12}^2=B\mu$. The stau block is  
given by 
\beq
\label{eq:subtautau}
\hskip -4mm
{\bold M_{\tilde\tau\tilde\tau}^2}\!=\!\! 
\left[\matrix{ \hskip -0.5mm
\half h_{\tau}^2v_1^2\!-\!\quarter g^2(v_1^2\!-\!v_2^2)\!+\!\mu\epsilon_3 
{{v_1}\over{v_3}}\!-\!B_2\epsilon_3{{v_2}\over{v_3}}\!+\!{{t_3}\over{v_3}} 
& {1\over{\sqrt{2}}}h_{\tau}(A_{\tau}v_1-\mu v_2) \cr 
\vb{20}
{1\over{\sqrt{2}}}h_{\tau}(A_{\tau}v_1-\mu v_2) 
&\hskip -4mm m_{R_3}^2\!\!+\!\half h_{\tau}^2(v_1^2\!+\!v_3^2) 
\!-\!\quarter g'^2(v_1^2\!-\!v_2^2\!+\!v_3^2) }
\hskip -1mm \right] 
\eeq
We recover the usual stau mass matrix again by replacing  
$v_3=\epsilon_3=0$, nevertheless, we need to replace the expression of the 
third tadpole in Eq.~(\ref{eq:tadpoles}) before taking the limit. 
The mixing between the charged Higgs sector and the stau sector is 
given by the following $2\times2$ block: 
\begin{equation} 
{\bold M_{H\tilde\tau}^2}=\left[\matrix{ 
-\mu\epsilon_3-\half h_{\tau}^2v_1v_3+\quarter g^2v_1v_3 
& -B_2\epsilon_3+\quarter g^2v_2v_3 \cr 
\vb{20}
-{1\over{\sqrt{2}}}h_{\tau}(\epsilon_3v_2+A_{\tau}v_3) 
& -{1\over{\sqrt{2}}}h_{\tau}(\mu v_3+\epsilon_3v_1) 
}\right] 
\label{eq:subHtau} 
\end{equation} 
and as expected, this mixing vanishes in the limit $v_3=\epsilon_3=0$. 
The charged scalar mass matrix in Eq.~(\ref{eq:subdivM}), 
after setting $t_1=t_2=t_3=0$, has determinant 
equal to zero since one of the eigenvectors corresponds to the charged 
Goldstone boson with zero eigenvalue.

\begin{figure}[ht]
\begin{center}
\begin{tabular}{cc}
\includegraphics[bb= 80 100 525 700,height=75mm,angle=90]{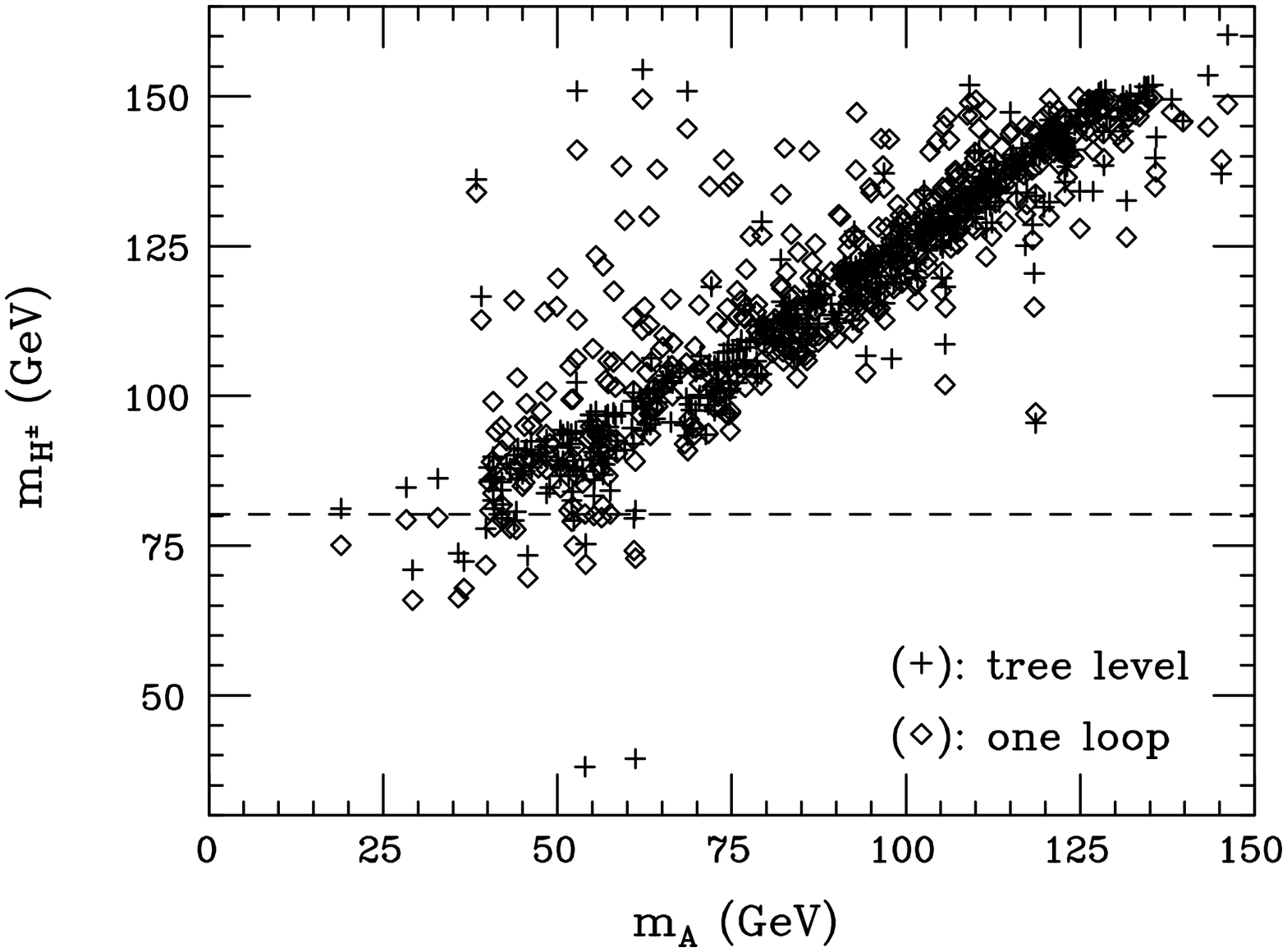}
&
\includegraphics[bb= 80 75 525 675,height=75mm,angle=90]{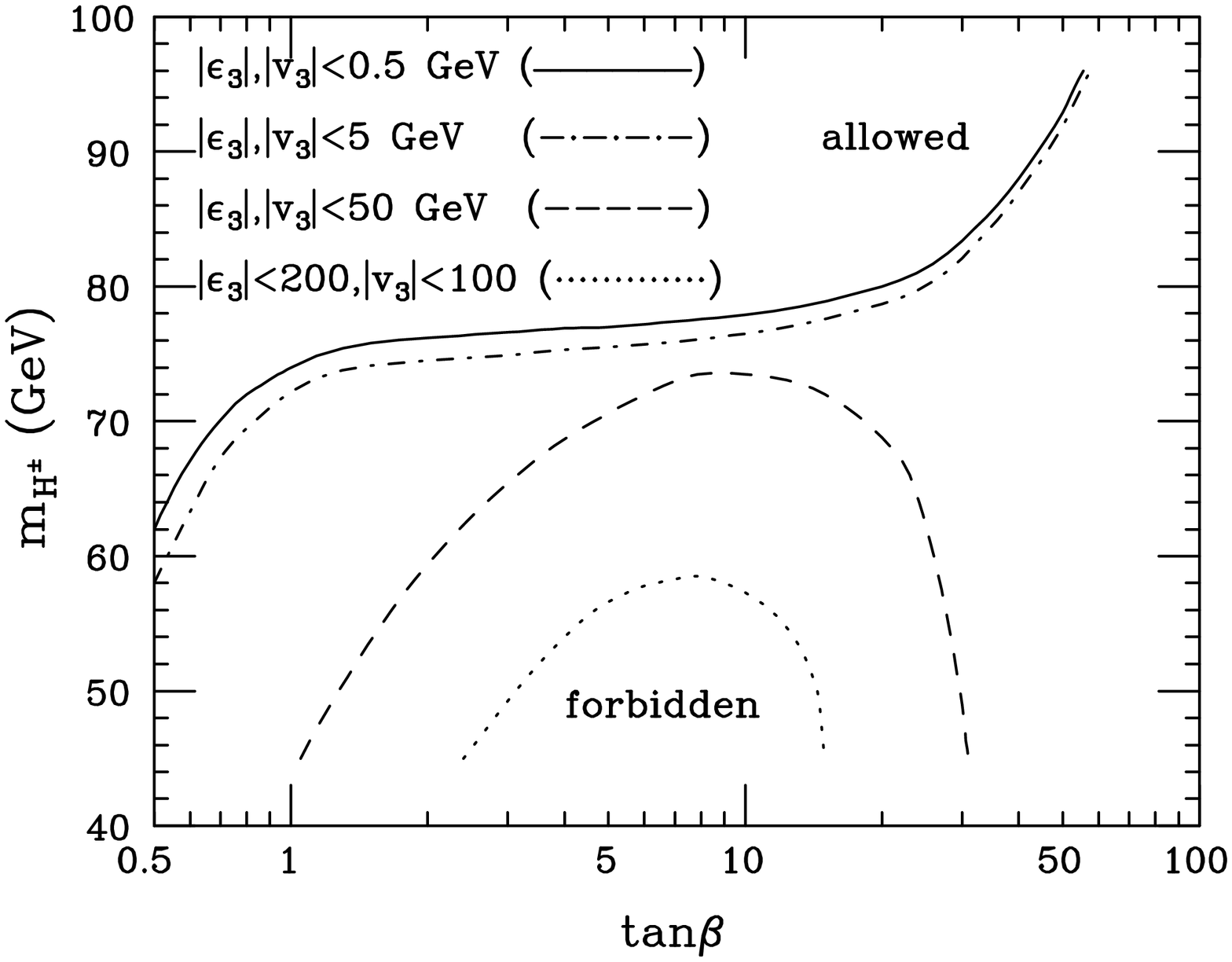}\cr
a)&b)
\end{tabular}
\end{center}
\vspace{-5mm}
\caption{\small 
a) Tree level and one--loop charged Higgs boson mass  
as a function of the CP--odd Higgs mass $m_A$. The horizontal 
dashed line corresponds to the $W$-boson mass.
b) Minimum of the charged Higgs boson mass versus $\tan\beta$.
Each curve corresponds to a different range of variation of the
R--parity violating parameters $ \epsilon_{3} $ and $ v_{3}$.
}
\label{fig:mchma} 
\end{figure}

The numerical study of the lowest-lying charged scalar boson mass has 
been done in ref.~\cite{charscalars}. The results are illustrated in 
Fig.~\ref{fig:mchma}a.
The main point to note is that $m_{H^{\pm}}$ can be 
lower than expected in the MSSM, even before including radiative 
corrections. This is due to negative contributions arising from the 
R--parity violating stau-Higgs mixing, controlled by the parameter 
$\epsilon_3$. An alternative way to display the influence of 
$\epsilon_3$ parameter 
on the charged Higgs boson mass can be seen in Fig.~\ref{fig:mchma}b.
In this figure the curves corresponding to different $\epsilon_3$ and
$v_3$ values delimit the minimum theoretically allowed charged Higgs
boson mass corresponding to those specific values.  

\begin{figure}[ht]
\begin{center}
\begin{tabular}{cc}
\includegraphics[bb= 80 100 525 700,height=75mm,angle=90]{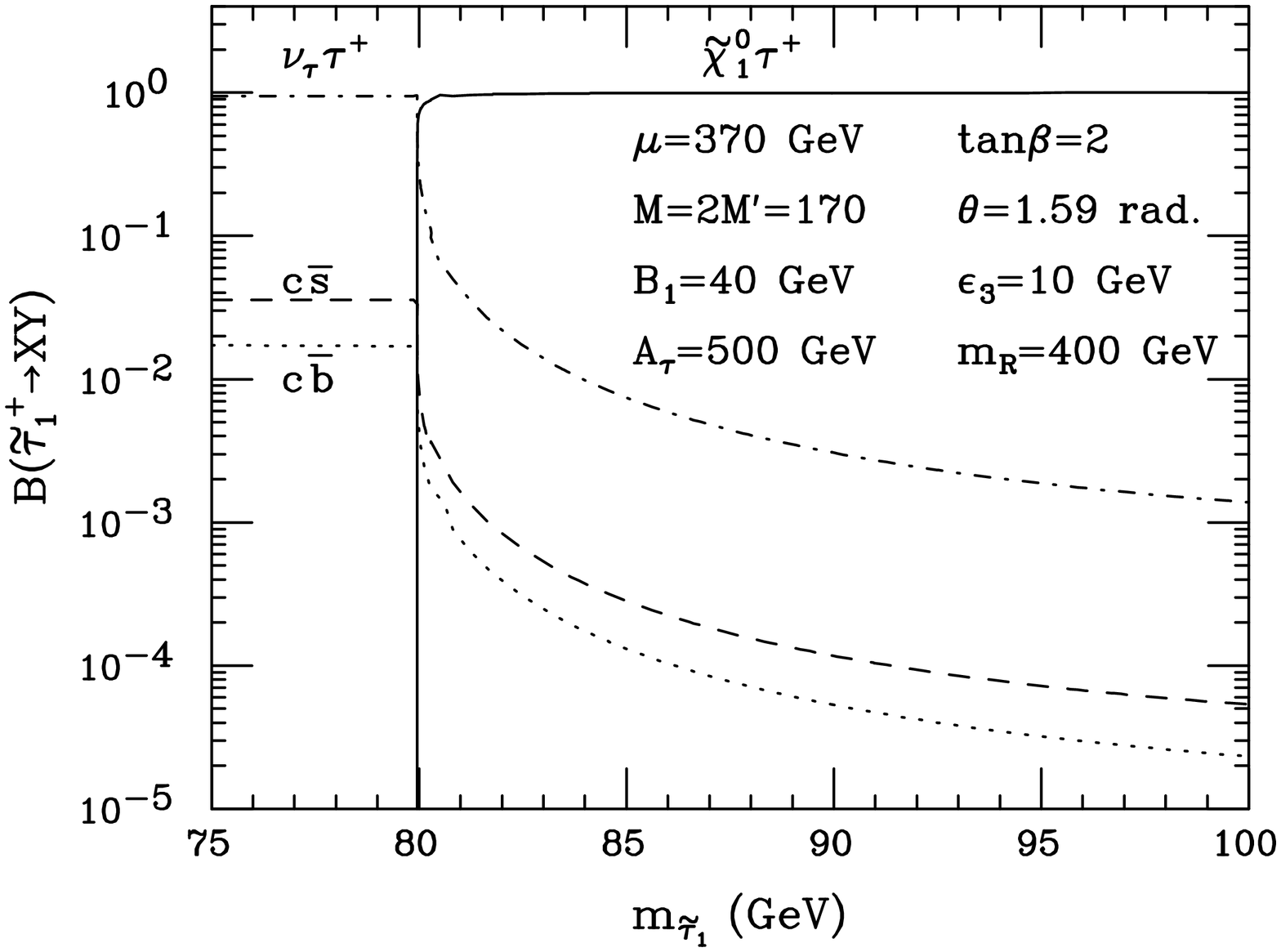}
&
\includegraphics[bb= 80 75 525 675,height=75mm,angle=90]{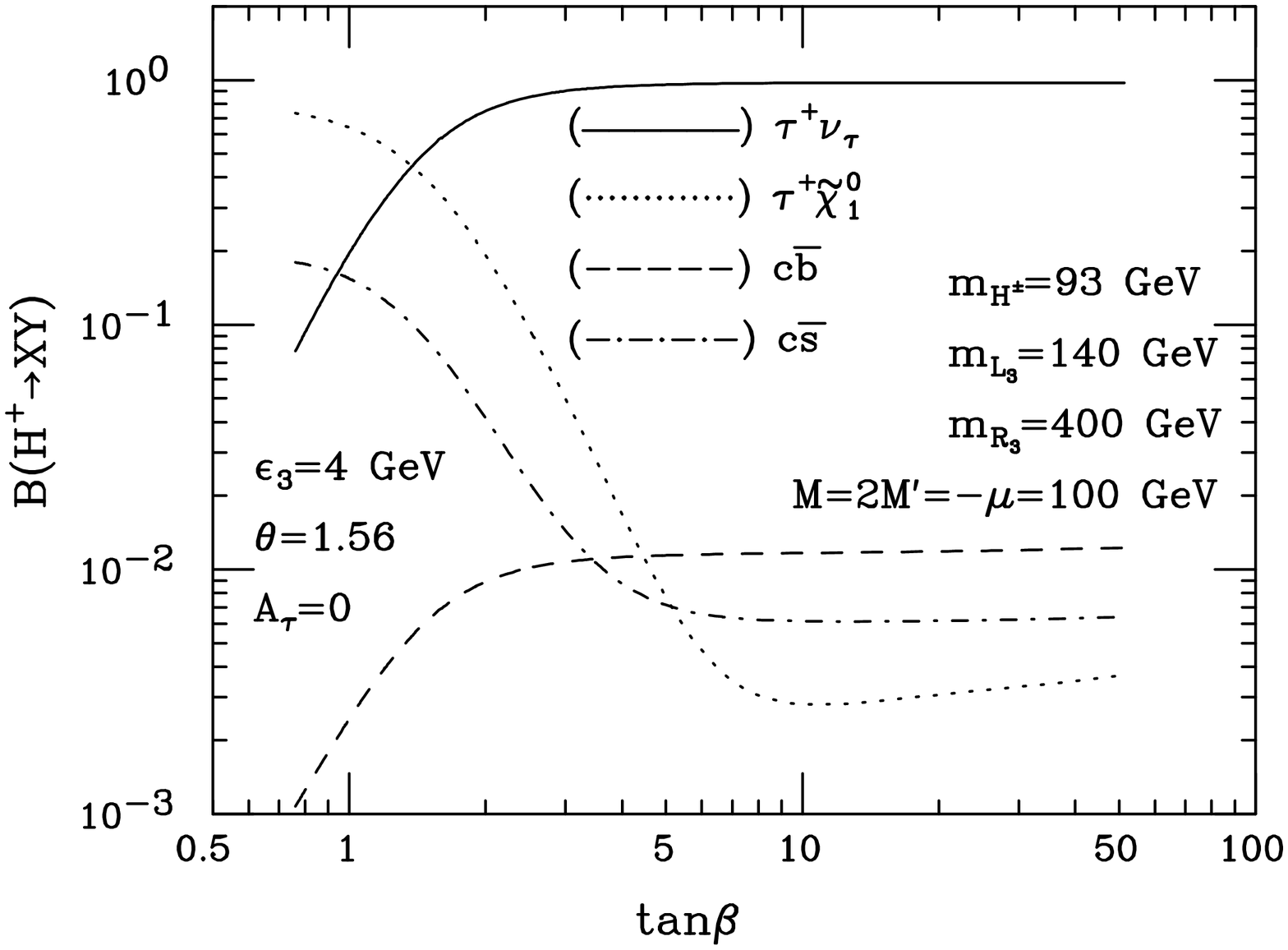}\cr
a)&b)
\end{tabular}
\end{center}
\vspace{-5mm}
\caption{\small
a) Stau branching ratios possible in our model for a particular choice
of parameters. Note the neutralino threshold below which only
R--parity violating decays are present.
b)
Charged Higgs branching ratios possible in our model for a
particular choice of parameters.
}
\label{fig:charBrs}
\end{figure}

\begin{figure}[ht]
\begin{center}
\includegraphics[bb= 80 100 525 700,height=75mm,angle=90]{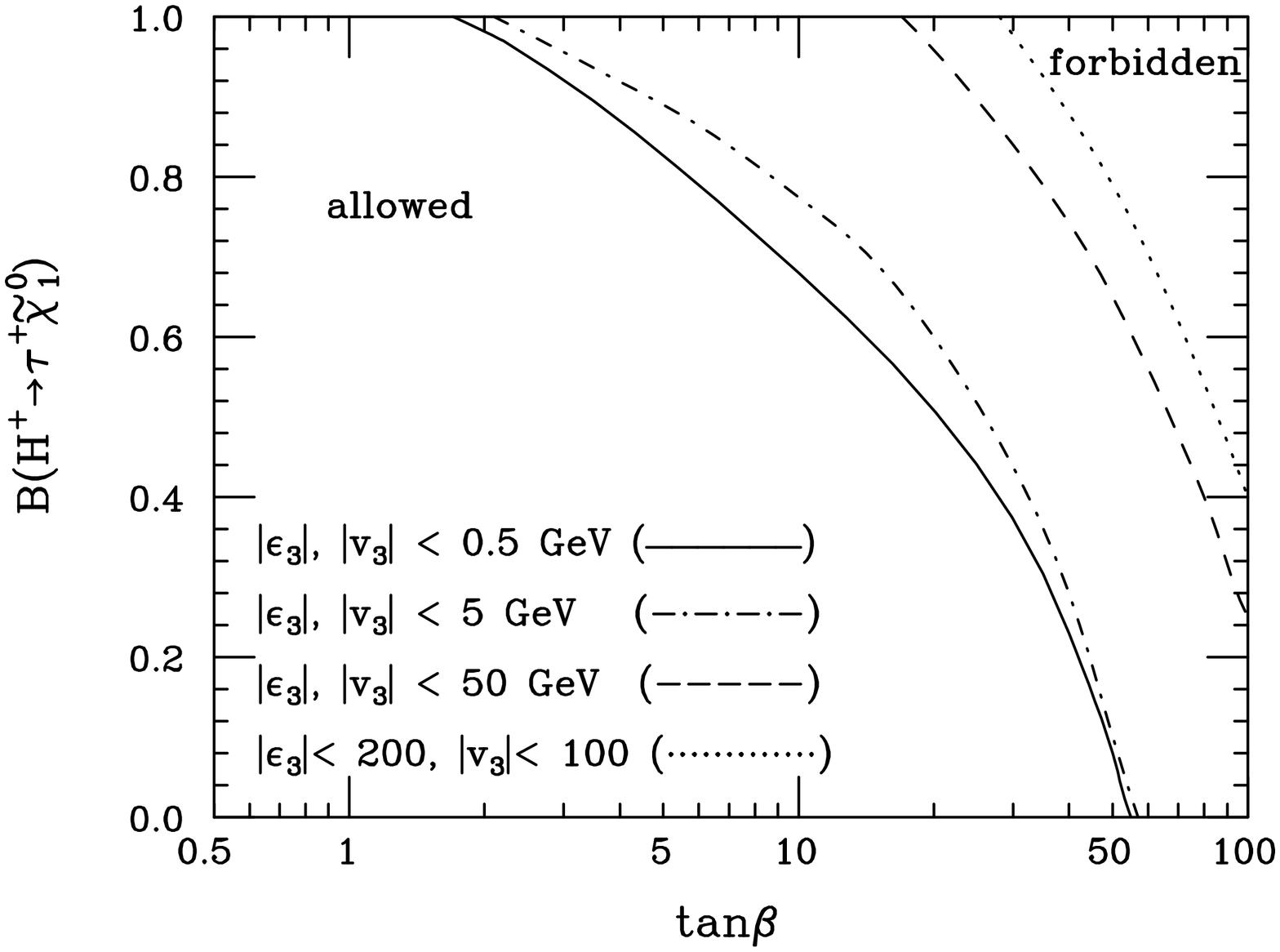}
\end{center} 
\vspace{-5mm}
\caption{\small
The curves denote the maximum attainable $R_p$-violating
charged Higgs branching ratio versus $\tan\beta$.
}
\label{rpv+br} 
\end{figure}

We now turn to a discussion of the charged scalar boson decays.  
In Fig.~\ref{fig:charBrs}a we display~\cite{charscalars} 
the stau decay branching ratios below and past the neutralino 
threshold and in Fig.~\ref{fig:charBrs}b the charged Higgs branching
ratios possible in the model for a particular set of chosen parameters.
Finally, for the case of the R--parity violating charged Higgs boson 
decays one can see from Fig.~\ref{fig:charBrs}b that the branching ratios into 
supersymmetric channels can be comparable or even bigger than the 
R--parity conserving ones, even for relatively small values of 
$\epsilon$ and $v_3$.  
Another way to see that the dominance of R--parity-violating Higgs
boson decays is not an accident of the above parameter choice is
illustrated in Fig.~\ref{rpv+br}. The various curves denote the
maximum attainable values for the R--parity-violating Higgs boson
branching ratio $ B(H^{+} \rightarrow \tau^{+}
\tilde\chi^{0}_{1})$. 

\subsection{Radiative Breaking}

In the previous discussion of the $\epsilon$--model the parameters
were varied at the weak scale with no restrictions besides the
experimental constraints on the masses of the particles. However, as
we have seen with the MSSM, the parameter space can be constrained if
we embed the theory in a grand unified scenario. This can also be done
in the $\epsilon$--model, both with~\cite{epsrad} and 
without~\cite{btepsrad} $b-\tau$ unification. We will describe below
these two possibilities.

\subsubsection{Radiative Breaking in the $\epsilon$ model: The minimal case}

At $Q = M_{GUT}$ we assume the standard minimal supergravity
unifications assumptions, 
\bea
&&A_t = A_b = A_{\tau} \equiv A \ ; \
B=B_2=A-1 \:, \cr
&&\vb{16}
m_{H_1}^2 = m_{H_2}^2 = M_{L}^2 = M_{R}^2 = 
M_{Q}^2 =M_{U}^2 = M_{D}^2 = m_0^2 \:, \cr
&&\vb{16}
M_3 = M_2 = M_1 = M_{1/2} 
\eea
In order to determine the values of the Yukawa couplings and of the
soft breaking scalar masses at low energies we first run the RGE's from
the unification scale $M_{GUT} \sim 10^{16}$ GeV down to the weak
scale. 
We randomly give values at the unification scale
for the parameters of the theory. 
\beq
\begin{array}{ccccc}
10^{-2} & \leq &{h^2_t}_{GUT} / 4\pi & \leq&1 \cr
\vb{14}
10^{-5} & \leq &{h^2_b}_{GUT} / 4\pi & \leq&1 \cr
\vb{14}
-3&\leq&A/m_0&\leq&3 \cr
\vb{14}
0&\leq&\mu^2_{GUT}/m_0^2&\leq&10 \cr
\vb{14}
0&\leq&M_{1/2}/m_0&\leq&5 \cr
\vb{14}
10^{-2} &\leq& {\epsilon^2_3}_{GUT}/m_0^2 &\leq& 10\cr 
\end{array}
\eeq
The value of ${h^2_{\tau}}_{GUT}/ 4 \pi$ is defined in such a way
that we get the $\tau$ mass correctly. 
As the charginos mix with the tau
lepton, through a mass matrix is given by
\beq
{\bf M_C}=\left[\matrix{ 
M & {\textstyle{1\over{\sqrt{2}}}}gv_u & 0 \cr 
\vb{18}
{\textstyle{1\over{\sqrt{2}}}}gv_d & \mu &  
-{\textstyle{1\over{\sqrt{2}}}}h_{\tau}v_3 \cr 
\vb{18}
{\textstyle{1\over{\sqrt{2}}}}gv_3 & -\epsilon_3 & 
{\textstyle{1\over{\sqrt{2}}}}h_{\tau}v_d}
\right] 
\eeq
Imposing that one of
the eigenvalues reproduces the observed tau mass $m_{\tau}$, $h_{\tau}$
can be solved exactly as~\cite{epsrad}
\beq
h_{\tau}^2={{2m_{\tau}^2}\over{v_d}}\left[
{{1+\delta_1}\over{1+\delta_2}}
\right]
\eeq
where the $\delta_i\,$, $i=1,2$, depend on $m_{\tau}$, on the SUSY
parameters $M,\mu,\tan\beta$ and on the $R_p$ violating parameters
$\epsilon_3$ and $v_3$. 
It can be shown that~\cite{epsrad}
\beq
\lim_{\epsilon_3 \ra 0} \delta_i = 0
\eeq

After running the RGE we have a 
complete set of parameters, Yukawa couplings and soft-breaking masses 
$m^2_i(RGE)$ to study the minimization. This is done by using a method
similar to the one described before in Section 2:

\begin{enumerate}

\item
We start with random values for $h_t$ and $h_b$ at $M_{GUT}$. 
The value of $h_{\tau}$ at $M_{GUT}$
is fixed in order to get the correct $\tau$ mass.

\item
The value of $v_1$ is determined from $m_{b}=h_b v_1/ \sqrt{2}$ for
$m_{b}=2.8$ GeV (running $b$ mass at $m_Z$). 

\item
The value of $v_2$ is determined from $m_{t}=h_t v_2/ \sqrt{2}$ for
$m_{t}=176 \pm 5$ GeV. If 
\beq
v_1^2+v_2^2 > v^2=\frac{4}{g^2}\, m^2_W = (246 \hbox{ GeV})^2
\eeq
we go back and choose another starting point.

\item
The value of $v_3$ is then obtained from
\beq
v_3=\pm\, \sqrt{\frac{4}{g^2}\, m^2_W -v_1^2 -v_2^2}
\eeq

\end{enumerate}

We see that the freedom in $h_{t}$ and $h_{b}$ at $M_{GUT}$ can be
translated into the freedom in the mixing angles $\beta$ and
$\theta$. Comparing, at this point, with the MSSM we have one extra
parameter $\theta$. We will discuss this in more detail below. In 
the MSSM we would have $\theta=\pi/2$.

After doing this, for each point in parameter space, we solve the extremum
equations, for the soft breaking
masses, which we now call $m^2_i$ ($i=H_1,H_2,L$). 
Then we calculate numerically the eigenvalues for the real and
imaginary part of the neutral scalar mass-squared matrix. If they are
all positive, except for the Goldstone boson, the point is a good one. 
If not, we go back to the next random value. 
As before, we end up
with a set of solutions for which
the $m^2_i$ obtained from the minimization
of the potential differ from those obtained from the RGE, which we
call  $m^2_i(RGE)$. 
Our  goal is to find solutions that obey
\beq
m^2_i=m^2_i(RGE) \quad \forall i
\eeq
To do that we define a function
\beq
\eta= max \left( \frac{m^2_i}{m^2_i(RGE)},\frac{m^2_i(RGE)}{m^2_i}
\right) \quad \forall i 
\eeq
that satisfies $\eta \ge 1$. Then we are all set for a minimization
program. For this we used the CERN Library Program {\tt MINUIT}.
Following this procedure we were able to find~\cite{epsrad} plenty
of solutions.

Let us discuss the counting of free
parameters in this model and in the minimal N=1 supergravity unified
version of the MSSM. In the MSSM we have the parameters shown in
Table~\ref{table:SUGRA}. Normally the two extra parameters are taken to
be the masses of the Higgs bosons $h$ and $A$, the lightest CP-even
and the CP-odd states, respectively.
For the $\epsilon$--model the situation is described in
Table~\ref{table:epsilon}. As we have said before there is an extra
parameter. 
Finally, we note that in either case, the sign of the mixing parameter
$\mu$ is physical and has to be taken into account.

\begin{table}[ht]
\begin{center}
\begin{tabular}{|c|c|c|}\hline
Parameters & Conditions & Free Parameters \cr \hline\hline
$h_t$, $h_b$, $h_{\tau}$, $v_1$, $v_2$, $v_3$
&$m_W$, $m_t$, $m_b$, $m_{\tau}$ & $\tan\beta$, $\cos \theta$ \cr \hline
$A$, $m_0$, $M_{1/2}$, $\mu$, $\epsilon_3$
&$t_i=0$, $i=1,2,3$& 2 Extra free param. \cr \hline
Total = 11&Total = 7 &Total = 4\cr\hline
\end{tabular}
\end{center}
\caption{Counting of free parameters in the $\epsilon$--model}
\label{table:epsilon}
\end{table}

\subsubsection{Gauge and Yukawa Unification in the $\epsilon$ model}

Besides achieving gauge coupling unification,
GUT theories also reduce the number of free parameters in the Yukawa
sector.  
In $SU(5)$ models, $h_b=h_{\tau}$ at $M_{GUT}$. The
predicted ratio $m_b/m_{\tau}$ at $M_{WEAK}$ agrees with the
experimental values. 
In the MSSM a relation between $m_{top}$ and
$\tan\beta$ is predicted. Two solutions are possible: low and high 
$\tan\beta$ .
In $SO(10)$ and $E_6$ models $h_t=h_b=h_{\tau}$ at $M_{GUT}$.
In this case, only the large $\tan\beta$ solution survives.
Recent global fits of low energy data ($B(b\rightarrow s\gamma)$ and  
the lightest Higgs mass) to the MSSM
show that it is hard to reconcile these constraints
with the large $\tan\beta$ solution.  Also the low $\tan\beta$ solution
with $\mu<0$ is disfavored.

Motivated by these considerations we analyzed the gauge and Yukawa
unification in the $\epsilon$--model. We found~\cite{btepsrad} 
that the $\epsilon$--model allows $b-\tau$ Yukawa unification for
any value of $\tan\beta$ and satisfying perturbativity of the
couplings.  We also found the $t-b-\tau$ Yukawa unification 
easier to achieve than in the MSSM, occurring in a 
wider high $\tan\beta$ region. We will describe below how we got
these results.

As before $h_{\tau}$ can be solved exactly 
\beq
h_{\tau}^2={{2m_{\tau}^2}\over{v_d}}\left[
{{1+\delta_1}\over{1+\delta_2}}
\right]
\eeq
where the $\delta_i\,$, $i=1,2$, depend on $m_{\tau}$, on the SUSY
parameters $M,\mu,\tan\beta$ and on the $R_p$ violating parameters
$\epsilon_3$ and $v_3$.
Also $h_t $ and $h_b$ are related to $m_t$ and $m_b$
\beq
m_t = h_t \frac{v}{\sqrt2} \sin \beta \sin \theta\,, \: \: \: \: \:
m_b = h_b \frac{v}{\sqrt2} \cos \beta \sin \theta 
\eeq
where
\beq
v=2m_W/g \qquad \tan\beta=v_u/v_d \qquad \cos\theta=v_3/v
\eeq

In our approach we divide the evolution into three
ranges: {\it i)} From $m_{Z} \ra m_t$ we use running fermion masses and gauge 
couplings. {\it ii)} From $m_t \ra M_{SUSY}$ we use the
two-loop SM RGE's including the quartic Higgs coupling $\lambda$.
{\it iii)} Finally from $M_{SUSY} \ra M_{GUT}$ we use the two-loop
RGE's. 
Using a top $\ra$ bottom 
approach we randomly vary the unification scale
$M_{GUT}$ and the unified coupling $\alpha_{GUT}$ looking for
solutions compatible with the low energy data
\beqa
&&\alpha^{-1}_{em}(m_Z) = 128.896 \pm0.090\cr
&&\vb{16}
\sin^2\theta_w(m_Z) =
0.2322 \pm 0.0010\cr
&&\vb{16}
\alpha_s(m_Z)=0.118 \pm 0.003
\eeqa
We get a region centered around 
$M_{GUT} \approx
2.3 \times10^{16}$ GeV ${\alpha_{GUT}}^{-1} \approx 24.5$
Next we use a bottom $\ra$ top approach to 
study the unification of Yukawa couplings using two-loop
RGEs. We take 
\beqa
&&m_W = 80.41 \pm 0.09\ GeV\cr
&&\vb{16}
m_{\tau}=1777.0 \pm 0.3 \ MeV \cr
&&\vb{16}
m_b(m_b) = 4.1\ \hbox{to}\ 4.5\ GeV 
\eeqa
We calculate the running masses 
$m_{\tau}(m_t)=\eta_{\tau}^{-1}m_{\tau}(m_{\tau})$ and
$m_b(m_t)=\eta_b^{-1}m_b(m_b)$ where $\eta_{\tau}$ and $\eta_b$
include three--loop order QCD and one--loop order QED. 
At the scale $Q=m_t$ we keep as a free parameter the running top quark
mass $m_t(m_t)$ and vary randomly the SM quartic Higgs coupling
$\lambda$.  In doing the running we used the following boundary conditions:

\begin{enumerate}
\item
At scale $Q=m_t$
\beq
\lambda_i^2(m_t)=2m_i^2(m_t)/v^2 \quad ; \quad
i=t,b,\tau
\eeq
\item
At scale $Q=M_{SUSY}$
\beqa
&&\lambda_t(M_{SUSY}^-)=h_t (M_{SUSY}^+) \sin\beta\sin\theta \cr
&&\vb{16}
\lambda_b(M_{SUSY}^-)=h_b (M_{SUSY}^+) \cos\beta\sin\theta \cr
&&\vb{16}
\lambda_{\tau}(M_{SUSY}^-)=h_{\tau} (M_{SUSY}^+) \cos\beta\sin\theta
\sqrt{{1+\delta_2}\over{1+\delta_1}}
\eeqa
\end{enumerate}

where $h_i$ denote the Yukawa couplings of our model and $\lambda_i$
those of the SM.
The boundary condition for the quartic Higgs coupling is
\beqa
\lambda(M_{SUSY}^-) &=& \cr
&&\vb{18}\hskip -3cm
\frac{1}{4}
\Big[(g^2(M_{SUSY}^+)+g'^2(M_{SUSY}^+) \Big] (\cos2\beta\sin^2\theta+
\cos^2\theta)^2
\eeqa
The MSSM limit is obtained setting $\theta \to \pi/2$ i.e. $v_3=0$.

\begin{figure}[ht]
\begin{center}
\vspace{-10mm}
\includegraphics[height=8cm]{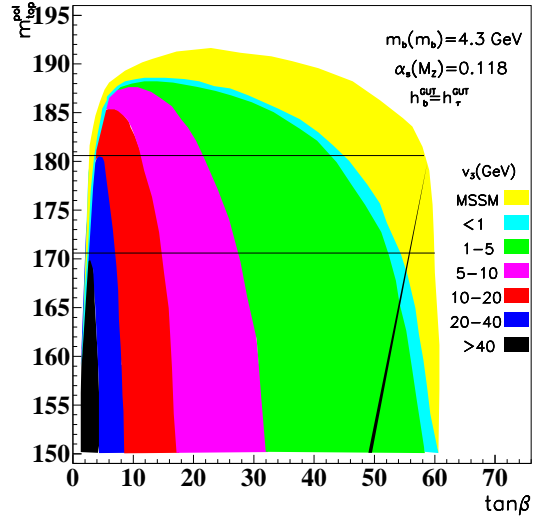}
\end{center}
\vspace{-10mm}
\caption{\small
Top quark mass as a function of $\tan\beta$ for different
values of the R--Parity violating parameter $v_3$. Bottom quark and
tau lepton Yukawa couplings are unified at $M_{GUT}$. The horizontal
lines correspond to the $1\sigma$ experimental $m_t$
determination. Points with $t-b-\tau$ unification lie in the diagonal
band at high $\tan\beta$ values. We have taken $M_{SUSY}=m_t$.
}
\label{fig:aretop}
\end{figure}

The results are summarized in Fig.~\ref{fig:aretop}. 
The dependence of our results on $\alpha_s$ and $m_b$
is totally analogous to what happens in
the MSSM. The upper bound on $\tan\beta$, which
is $\tan\beta\lsim 61$ for $\alpha_s=0.118$, increases with $\alpha_s$
and becomes $\tan\beta\lsim 63$ (59) for $\alpha_s=0.122$ (0.114).
The top mass value for which unification is achieved for any
$\tan\beta$ value within the perturbative region increases with
$\alpha_s$, as in the MSSM.  
As for the dependence on $m_b$, if we consider $m_b(m_b)=4.1$ (4.5)
GeV then the upper bound of this parameter is given by $\tan\beta\lsim
64$ (58). In addition, the MSSM region is narrower (wider) at high
$\tan\beta$ compared with the $m_b(m_b)=4.3$ GeV case.
The line at high $\tan\beta$ values corresponds
to points where $t-b-\tau$ unification is achieved. Since the region
with $|v_3|<5$ GeV overlaps with the MSSM region, it follows that
$t-b-\tau$ unification is possible in this model for values of $|v_3|$
up to about 5 GeV, instead of 50 GeV or so, which holds in the case of
bottom-tau unification.

\section*{Acknowledgments}
This work was supported in part by the TMR network 
grant ERBFMRX-CT960090 of the European Union.

\end{document}